\documentclass[aip,jcp,graphicx]{revtex4-1}
\usepackage{graphicx}
\usepackage{dcolumn}
\usepackage{bm}
\draft 
\usepackage{ gensymb }

\usepackage{xr} 
\usepackage[format=plain,justification=centerlast]{caption}

\begin{document}

\title{Reactive Atomistic Simulations of Diels-Alder Reactions: the Importance of Molecular Rotations} 



\affiliation{Department of Chemistry, University of Basel\\ Klingelbergstrasse 80, 4056 Basel, Switzerland }
\author{Ux\'ia Rivero}
\author{Oliver T. Unke}
\author{Markus Meuwly}
\email[]{m.meuwly@unibas.ch}
\author{Stefan Willitsch}
\email[]{stefan.willitsch@unibas.ch}



\date{\today}

\begin{abstract} 
The Diels-Alder reaction between 2,3-dibromo-1,3-butadiene and maleic
anhydride has been studied by means of multisurface adiabatic reactive molecular dynamics and the PhysNet neural network architecture. This system is used as a prototype to explore the concertedness, synchronicity and possible ways of promotion of Diels-Alder reactions. Analysis of the minimum dynamic path indicates that rotational energy is crucial ($\sim$~65\%) to drive the system towards the transition state in addition to collision energy ($\sim$~20 \%). Comparison with the reaction of butadiene and maleic anhydride shows that the presence of bromine substituents in the diene accentuates the importance of rotational excitation to promote the reaction. At the high total energies at which reactive events are recorded, the reaction is found to be direct and mostly synchronous.
\end{abstract}

%
%

\maketitle

\section{Introduction}
 	The regio- and stereoselective Diels-Alder reaction in which a diene reacts with a dienophile to form a cyclic product is widely used in synthetic organic chemistry.\cite{diels28a, ishihara14a} In this reaction, two $\rm \sigma$ bonds and one $\rm \pi$ bond are formed from three $\rm \pi$ bonds as depicted in Scheme 1. 
    \begin{figure*}[h]
    	\includegraphics[width=0.9\textwidth]{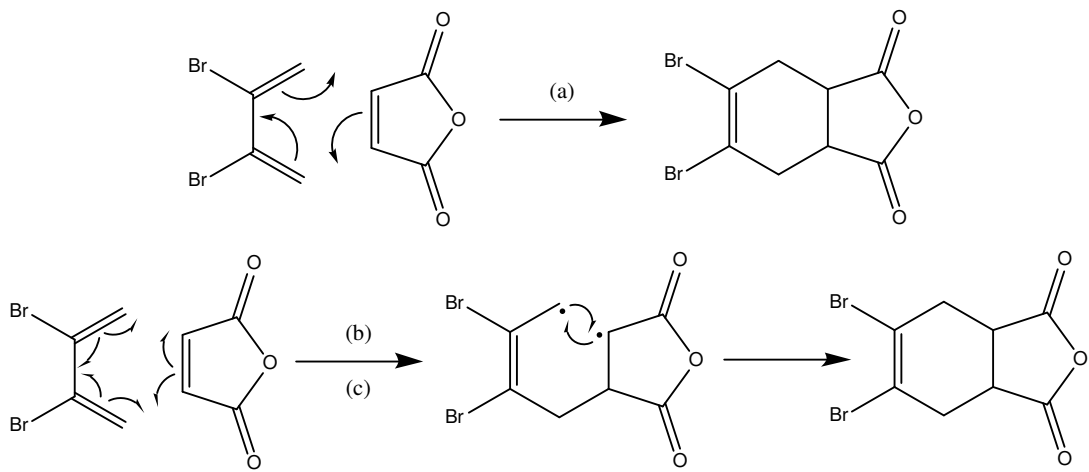}
    	\centering
    	\caption*{Scheme 1. Diels-Alder reaction between 2,3-dibromo-1,3-butadiene (DBB) and maleic anhydride (MA): (a) Concerted mechanism, (b) and (c) stepwise mechanism with a short-lived and a long-lived intermediate, respectively.}
    \end{figure*}

 	There have been many experimental and theoretical studies aimed at unveiling the Diels-Alder mechanism at a molecular level and its dependence on the geometric and electronic characteristics of the reactants.\cite{houk95a, yepes13a, souza16a, domingo14a, horn96a, diau99a, saettel02a, singleton01a, goldstein96a, sakai00a, domingo09a, donoghue06a} Since two bonds are formed, questions about concertedness and synchronicity render this reaction also interesting from a theoretical point of view. 
 	
 	The concertedness of a reaction is determined by the topology of the potential energy surface (PES).\cite{minkin99a} A concerted mechanism has only one transition state (TS) between reactants and products such that the reaction takes place in a single step. In a stepwise mechanism, there is at least one stable intermediate between reactants and products.
 	
 	The time elapsed between the formation of the first and the second bond defines the synchronicity.\cite{minkin99a} It is usually thought that symmetric TSs give rise to synchronous processes in which both bonds are formed at the same time while asymmetric TSs lead to asynchronous processes in which first one bond is formed and then the second one follows. This definition of synchronicity has recently been challenged by some authors who argue that it should not be defined from a geometric but from a dynamic point of view since the connection between spatial quantities and temporal concepts may not always hold.\cite{souza16a, black12a}
 	
 	There is a long-standing discussion about synchronicity and concertedness in the context of Diels-Alder reactions.\cite{houk95a} While it is often believed to be a concerted, synchronous reaction involving an aromatic TS governed by the Woodward-Hoffmann rules,\cite{hoffmann68a} experiments and calculations show that this is not true for all cases.\cite{horn96a, diau99a, saettel02a, singleton01a, goldstein96a, wilsey99a, hofmann99a,bouchoux04a} In principle, one can think of three possible mechanisms (see Scheme~1): (a) concerted, (b) stepwise with a short-lived intermediate, and (c) stepwise with a long-lived intermediate whose lifetime allows for the rotation around a C-C bond. Note that when the system cannot rotate around a C-C bond, as in cases (a) and (b), the reaction is stereo-selective and only the s-\textit{cis} conformer of the diene will yield the cyclic Diels-Alder product. Mechanism (c), on the other hand, is not stereo-selective and the s-\textit{trans} conformer of the diene could also participate in the reaction.
 	
 	Computational studies of Diels-Alder reactions are especially sensitive to the choice of method and basis set. \cite{gayatri11a, linder13a} Different electronic structure methods give different TS structures and activation energies as was summarized in the introduction of our previous paper (Ref.~\onlinecite{rivero17a}). This leads to different conclusions regarding the synchronicity and concertedness of the reaction. In general, neutral reactions occur in a concerted fashion while in cationic systems a stepwise mechanism is favored. \cite{saettel02a, singleton01a, goldstein96a} However, the border between these two mechanisms is diffuse. Moreover, some studies show that both concerted and stepwise mechanisms are present at the temperatures at which these reactions are usually performed due to the energetic proximity of both pathways in many systems making dynamical studies crucial for the exploration of these reactions. To the best of our knowledge, all theoretical studies on the dynamics of Diels-Alder reactions have started from TS-like structures\cite{souza16a, black12a, tan18a, wang09a, liu16a} or have used steered dynamics to drive the reaction.\cite{soto16a} These procedures will most likely bias the final result and do not allow the direct calculation of reaction rates. Gas-phase reactive molecular dynamics simulations starting from an equilibrated ensemble of a statistically significant number of initial conditions, on the other hand, have recently been shown to provide molecular-level details into reactions relevant to atmospheric chemistry \cite{meuwly19a, brickel17a} and reactions in the hypersonic regime.\cite{koner18a, alpizar17a} 
 	
 	From an experimental perspective, the most precise data on reaction mechanisms and dynamics can be gained from gas-phase studies performed under single-collision conditions. As the progress in molecular-beam experiments allows the probing of ever-larger systems under precisely defined conditions,\cite{chang13a, willitsch17a} the open questions pertaining to the mechanistic details of Diels-Alder reactions become an attractive target of study. Since only scarce information is available on these important mechanistic aspects,\cite{eberlin93a} we present here a molecular dynamics study of the Diels-Alder reaction between 2,3-dibromo-1,3-butadiene (DBB) and maleic anhydride (MA) (see Scheme~1) which may serve as a guide for future experiments.
 	
 	We have chosen DBB because it is a generic diene which fulfills the experimental requirements for conformational separation of its isomers by electrostatic deflection of a molecular beam,\cite{chang13a, willitsch17a} thus enabling the characterization of conformational aspects and specificities of the reaction. MA is a widely used, activated dienophile which due to its symmetry simplifies the possible products of the reaction. The reaction of DBB and MA thus serves as a prototypical system well suited for the exploration of general mechanistic aspects of Diels-Alder processes in the gas phase.
 	
 	In this paper, we use reactive molecular dynamics simulations starting from the two reactant molecules approaching each other in order to simulate a collision experiment and address questions such as whether the reaction is synchronous, whether the mechanism is complex-mediated and how the reaction could be promoted.

\section{Methods}
    
    \subsection{Molecular Dynamics Simulations}
    \label{md}
    Atomistic simulations were carried out either with the CHARMM program\cite{CHARMM} using an initial parametrization from SwissParam\cite{zoete11a} or with the Atomic Simulation Environment (ASE)\cite{larsen17a} using the PhysNet neural network architecture designed for predicting energies, forces and dipole moments of chemical systems.\cite{unke2019a}
      All bonds involving hydrogen atoms were flexible and the time step used in the simulations was $\Delta t$~=~0.1~fs to ensure conservation of total energy. The velocity Verlet algorithm was used for the propagation of the equations of motion.\cite{swope82a}

	The initial SwissParam parametrization was modified in order to construct a multisurface adiabatic reactive molecular dynamics (MS-ARMD) \cite{nagy14a} force field for the present Diels-Alder reaction. Ensembles of structures for the parametrization of the MS-ARMD model were generated with CHARMM as follows: the optimization of the system with the adopted Newton-Raphson method was followed by 50~ps of heating dynamics, 50~ps of equilibration at 500~K, 60~ps of cooling down to 300~K and free \textit{NVE} (microcanonical ensemble) dynamics. The temperature was only raised up to 400~K for parts of the parametrization of the reactant van der Waals complex to avoid dissociation.
	
	For the PhysNet parametrization, initial ensembles for the different fragments\cite{huang2017a} were generated using the PM7 level of theory implemented in MOPAC2016\cite{stewart2013a,mopac2016} and subsequently augmented using adaptive sampling.
	
	In order to generate the initial conditions for the collision simulations, ensembles of the individual molecules (MA and DBB) at different vibrational temperatures were generated using CHARMM as described above. Heating and equilibration temperatures were modified depending on the desired final vibrational temperature ($T_{\rm vib}$). The reactants were placed at an initial distance of 20~$\rm \AA$~(taking the center of mass of each molecule as reference) with a random relative orientation. In order to tune the collision energy ($E_{\rm coll}$), the atomic velocities along the collision axis were modified. The impact parameter ($b$) was uniformly sampled by displacing one of the molecules along a perpendicular axis. Rotational temperature ($T_{\rm rot}$) was added following calculation of the moment of inertia tensor and assuming equipartition among the three rotational degrees of freedom. Excitation of specific vibrational normal modes was achieved by projecting the initial velocities onto the space of normal modes and modifying the kinetic energy of the desired normal mode. 
	
	The trajectories were considered reactive when the C-C distance between the carbon atoms involved in the two new bonds was smaller than 1.6~$\rm \AA$. The reactive cross section $\sigma$ was calculated as 
		\begin{equation}
		\label{impact-parameter}
		\sigma = 2 \pi b_{\rm max} \frac{1}{N_{\rm tot}} \sum_{i=1}^{N_{\rm reac}} b_i
		\end{equation}
	where $b_{\rm max}$ is the maximum impact parameter at which no reactive events are observed any more, $N_{\rm tot}$ the total number of trajectories, $N_{\rm reac}$ the number of reactive trajectories and $b_i$ the impact parameter of the reactive trajectory $i$. The reaction rate $k$ can be calculated from 
		\begin{equation}
		k = \sigma \cdot v_{\rm rel}
		\end{equation}
	where $v_{\rm rel}$ is the relative center of mass velocity of the colliding molecules.
	
	The total kinetic energy of the minimum-dynamic-path trajectories was decomposed either into the translational energy of the center of mass of the reactant molecules, the rotational energy corresponding to their angular momentum and vibrational energy as stated in equations~S1~-~S6 of the supplementary information or by projecting it onto the degrees of freedom of the system. The reference degrees of freedom were calculated as the eigenvectors of the Hessian matrix of the isolated, reactant molecules with geometries corresponding to the last point of each trajectory.
	
	\subsection{Parametrization of MS-ARMD}
	 The force fields for the reactant and product states were parametrized to reproduce DFT reference energies at the M06-2X/6-31G* level of theory which was found to yield the best accuracy at the DFT level for this type of reaction by Linder and Brinck\cite{linder13a, zhao08a} and following our previous work.\cite{rivero17a} All single point calculations were performed using Gaussian09.\cite{g09} Starting from parameters retrieved from SwissParam,\cite{zoete11a} first ensembles for reactants and products were generated as described in section~\ref{md}. Reference energies were calculated and the individual force fields were fitted using a downhill simplex algorithm.\cite{nelder65a} The standard CHARMM harmonic potentials for the description of C-C, C=C, C-O and C-Br bonds were replaced by Morse potentials (C-H and C=O bonds were kept as harmonic). Furthermore, in the reactant force field, the Lennard-Jones potentials between the four carbon atoms involved in the Diels-Alder reaction as well as between Br/O atoms were replaced by a generalized Lennard-Jones potential.\cite{nagy14a} The remaining terms were parametrized as in the standard CHARMM force field. 
	 
	 With the initial parametrization of the individual force fields, another ensemble of structures was generated for reactants and products with the new parameters and added to the training structures. The reference energies were calculated and the parametrization was further refined. This iterative procedure continued until the root-mean-square deviation (RMSD) of the newly generated ensemble was approximately the same as that of the previous iteration of the parametrization. For the reactant force field, 401 and 2064 structures where required for the parametrization of MA and DBB, respectively. The non-bonded terms of the van der Waals complex were parametrized with 2783 additional structures. For the parametrization of the product force field, 2589 structures were required. 
	 
	 The crossing region between the two force fields was smoothed following the internal reaction coordinate (IRC) by combining the final force fields for reactants and products with GAussian times POlynomial functions (GAPOs).\cite{nagy14a} A genetic algorithm was used for the fitting of the GAPOs. The global PES is given by Eq.~\ref{ms-armd} where $V_i(x)$ are the individual force fields, $w_i(x)$ their weights  and $\Delta V^{ij}_{{\rm GAPO},k} (x)$ the GAPOs.\cite{nagy14a} The product of the Diels-Alder reaction has two possible connectivities (see Fig.~S3 of the supplementary information). In order to make the parametrization of the product permutation invariant, two force fields that describe these two possible connectivities were used.
	\begin{equation}  
	V_{\rm MS-ARMD} = \sum\limits_{i=1}^{n} w_i(x) V_i(x) +
	 \sum\limits_{i=1}^{n-1} \sum\limits_{j=i+1}^{n} [w_i(x)+w_j(x)] 
		\sum\limits_{k=1}^{n_{ij}}\Delta V^{ij}_{{\rm GAPO},k} (x)
		\label{ms-armd} 
	\end{equation}
	
	\subsection{Parametrization of PhysNet}
	Reference data for training PhysNet (energies, forces and dipole moments) were calculated at the DFT M06-2X/6-31G* level of theory using Gaussian09.\cite{g09} All possible ``amon'' fragments\cite{huang2017a} for DBB, MA, and their reaction product were generated (378 in total) and different geometries for all fragments were sampled by running Langevin dynamics at 1000~K at the PM7 level of theory. After training PhysNet models on this initial dataset, additional structures were generated by adaptive sampling:\cite{behler2014a, behler2015a} an ensemble of 4~PhysNet models was used to run Langevin dynamics at 1000~K and new \textit{ab initio} data was calculated for geometries for which the energy predictions between the different models differed by more than 0.5~kcal/mol. For further details on the adaptive sampling method, see Ref.~\onlinecite{unke2019a}. The dataset was iteratively augmented in this fashion until no significant deviations between the predictions of individual PhysNet models could be observed (the final dataset contains 224483 structures which is approximately 50 times larger than it was required for the MS-ARMD model). PhysNet was then trained on 200000 structures of the final dataset (with 20000 additional structures used for validation) by minimizing the mean absolute error between neural network predictions and the reference data using the procedure given in Ref.~\onlinecite{unke2019a} (all hyperparameters of the neural network architecture and the training procedure were set to the values recommended in Ref.~\onlinecite{unke2019a}). The global PES is given by 
	\begin{equation}
	V_{\rm PhysNet} = \sum\limits_{i=1}^{N} E_i + k_e\sum_{i=1}^{N}\sum_{j>i}^{N}\frac{q_{i}q_{j}}{r_{ij}}
	\label{physnet}
	\end{equation}
	where $N$ is the total number of atoms, $k_e$ is the Coulomb constant, $r_{ij}$ is the distance between atoms $i$ and $j$, and $E_i$ and $q_{i}$ are atomic energy contributions and partial charges (corrected to guarantee charge conservation)\cite{unke2019a} predicted by PhysNet. Here, the Coulomb potential is damped at short distances to avoid numerical problems (see Ref.~\onlinecite{unke2019a}). The PhysNet architecture guarantees that Eq.~\ref{physnet} is invariant with respect to translations, rotations and permutation of atoms sharing the same element type.\cite{unke2019a}	
	
\section{Results and Discussion}
  \subsection{Parametrization of the Reactive Force Fields}

	\begin{figure*}[h!]
		\includegraphics[width=0.75\textwidth]{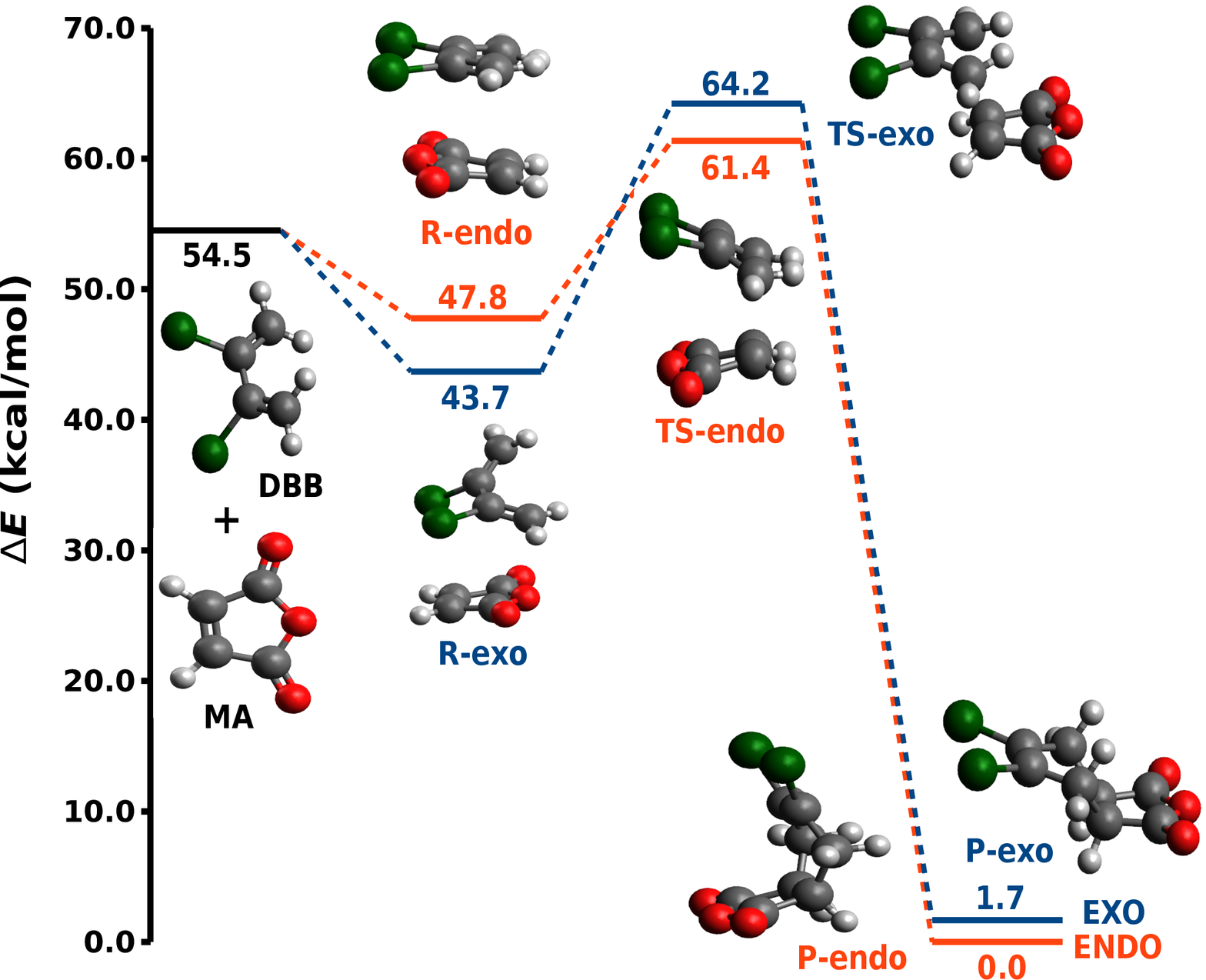}
		\caption[justification=justified]{Potential energy surface for the two possible Diels-Alder reaction paths (exo and endo in blue and orange, respectively) between dibromobutadiene (DBB) and maleic anhydride (MA) at the M06-2X/6-31G* level of theory. The relative energies in kcal/mol with respect to the endo product (P-endo) as well as the structures of minima and transition states are shown.}
		\label{neutpes}  
	\end{figure*}
	
	\begin{figure*}[h!]
		\includegraphics[width=0.95\textwidth]{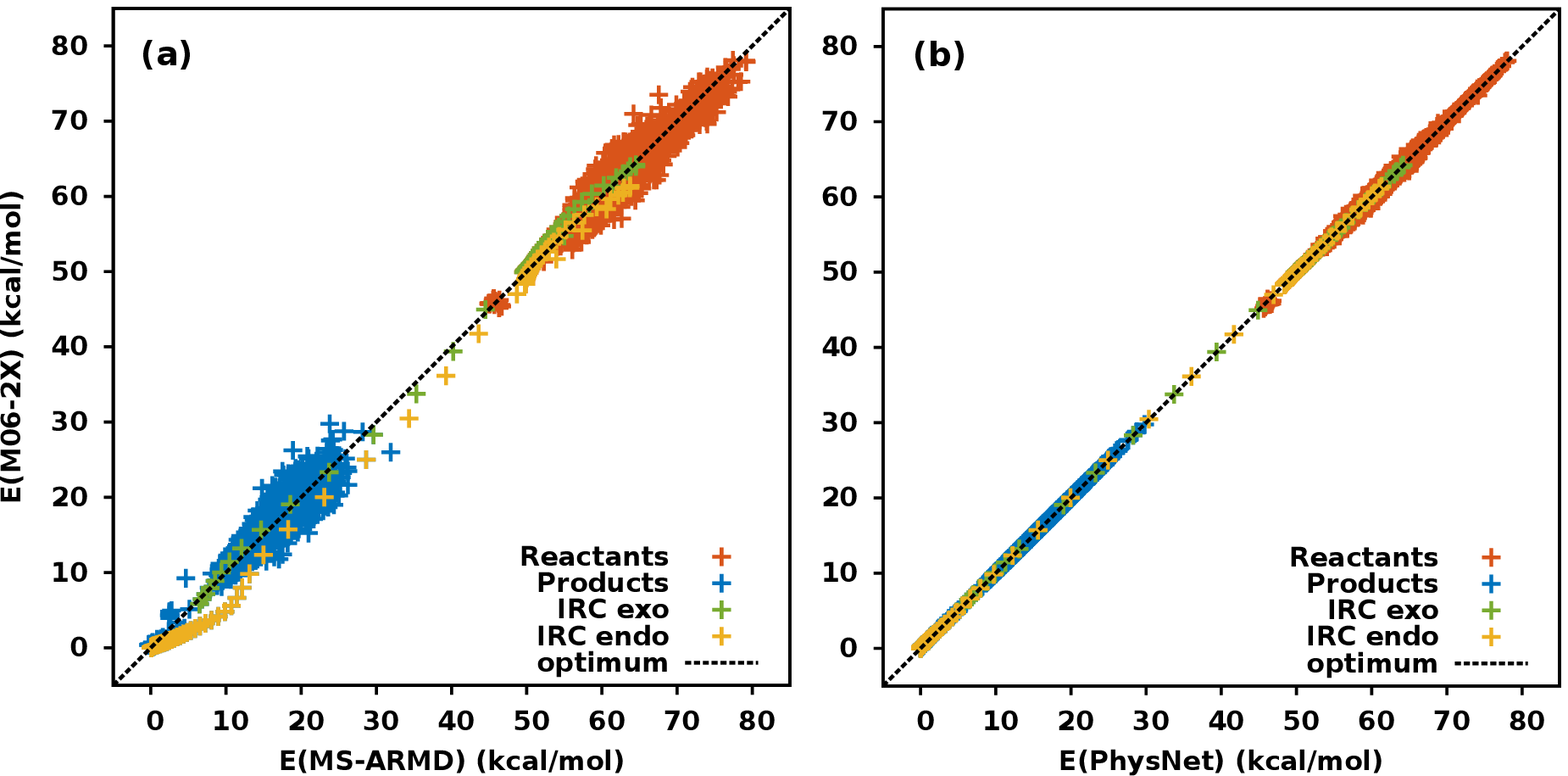}
		\caption{\label{parametrization}Energy correlation of 5512 M06-2X/6-31G* reference structures and (a) the MS-ARMD model with a total RMSD of 1.47~kcal/mol and $R^2$~=~0.9961 or (b) the PhysNet model with a total RMSD of 0.25~kcal/mol and $R^2$~=~0.9999.}		
	\end{figure*}

	Fig.~\ref{neutpes} shows stationary points on the PES for the Diels-Alder reaction between DBB and MA at the M06-2X/6-31G* level of theory. Note that DBB is in a \textit{gauche} conformation (the C=C$-$C=C dihedral angle has a value of 50\degree) since the s-\textit{cis} geometry is not a minimum on the PES. Due to the fact that both reactant molecules are symmetric, there are only two possible paths for the Diels-Alder reaction. These are referred to as ``endo'' and ``exo'' depending on the relative orientation of the reactants. The endo product (P-endo) is taken as the zero of the energy scale in Fig.~\ref{neutpes}. Normally, the intermolecular interactions favor the endo path which M06-2X correctly describes.\cite{alder36a, fernandez14a, rivero17a} The dissociation of the van der Waals complexes (R-endo and R-exo) towards the reactants is more favorable than the reaction over the barrier towards the Diels-Alder products (see Fig.~\ref{neutpes}). Judging from the PES and the geometries of the TSs, the reaction should be concerted and symmetric.   

	As discussed in the introduction, in order to answer questions such as whether the reaction is synchronous or complex mediated, the dynamics of the system must be studied. Therefore, we constructed a computationally efficient PES that allows us to run a statistically significant number of trajectories, such that diverse initial conditions can be sampled. The quality of the parametrized MS-ARMD model is shown in Fig.~\ref{parametrization}(a). Points generated during the parametrization of the product force field (2589 structures) and the parametrization of the non-bonded interactions of the reactant force field (2783) are shown, as well as the IRCs for the endo (81) and the exo (59) paths. The exo IRC was used for the parametrization of the GAPOs and thus it is better described by the model than the endo path. The energy of the endo IRC is slightly overestimated by the MS-ARMD model. The total RMSD is 1.47~kcal/mol over a range of 80~kcal/mol which we asses to be sufficient for an adequate characterization of the dynamics of the system. It is important to note that the s-\textit{trans} conformer has not been included and is not stable in this model. Thus, the reactants will not sample the entire conformational space leading to a possible overestimation of the reaction rates since the s-\textit{trans} conformer does not contribute to the reaction i.e. the number or unreactive trajectories increases when including the s-\textit{trans} conformer which leads to a reduction of the reaction flux and hence the rate. For the validation of the MS-ARMD model the reactant, transition state and product structures were minimized and compared to the DFT geometries (see Fig.~S1 of the supplementary information). The harmonic frequencies computed by DFT and the model were also compared (Fig.~S2 of the supplementary information). The parametrized model is given in Tables~SI-SVII of the supplementary information.
	
	For validation and direct comparison, a PhysNet model was parametrized for the same system. Fig.~\ref{parametrization}(b) reports on the quality of this PES by showing the correlation between the reference energies and the PhysNet predictions on the training structures of the MS-ARMD model with a total RMSD of 0.25~kcal/mol. The PhysNet model is significantly more accurate than MS-ARMD, but it has a computational cost 200 times higher.

  \subsection{Minimum Dynamic Path}

\begin{figure}[h!]
  			\includegraphics[width=0.95\textwidth]{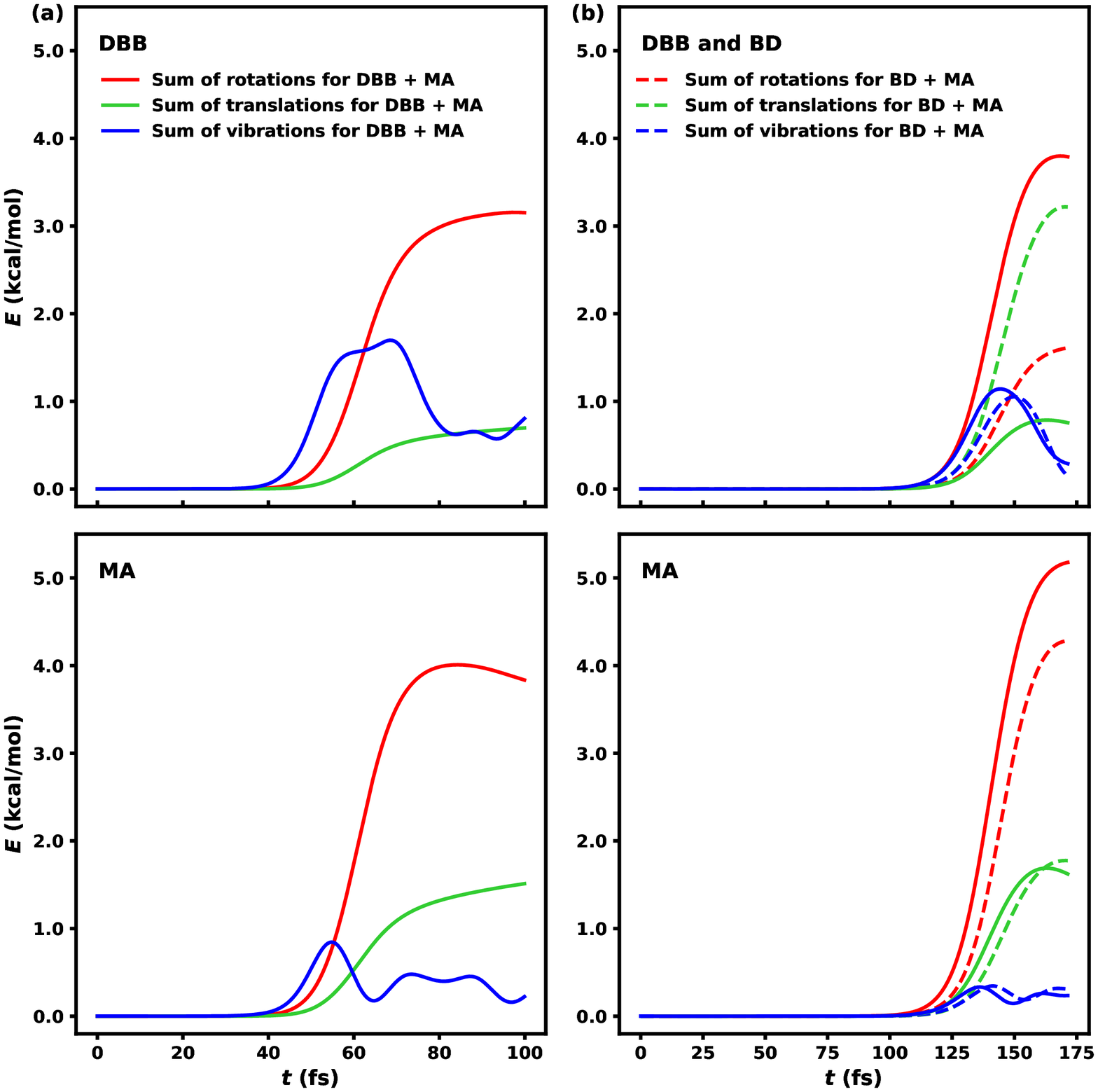}
  			\caption{Solid lines: projection of the total kinetic energy ($E$) onto the degrees of freedom (rotations, translations and vibrations) of dibromobutadiene (DBB) and maleic anhydride (MA) for the reaction of DBB~+~MA along the minimum dynamic path calculated with (a) the MS-ARMD model and (b) the PhysNet model. Dashed lines in panel (b): projection of the total kinetic energy onto the degrees of freedom of butadiene (BD) and MA for the reaction of BD~+~MA along the minimum dynamic path calculated with the PhysNet model. The trajectories start at the endo transition state and end at the reactants.}
  			\label{mdp}
  		\end{figure}

	A trajectory starting at the TS geometry without kinetic energy (the total energy of the trajectory is equal to the potential energy of the TS) follows	the minimum dynamic path (MDP).\cite{unke2019b} The projection of the total kinetic energy along the MDP towards the reactants onto the degrees of freedom of DBB and MA is shown in Fig.~\ref{mdp}(a) as sums of translational, rotational and vibrational energies. At $t$~=~0~fs, the system is at the TS and at $t$~=~100~fs it has arrived at the reactant state. At the beginning of the trajectory, the system contains no kinetic energy and moves only slowly, until at around $t$~=~50~fs more potential energy is converted into kinetic energy. By projecting the total kinetic energy onto the different degrees of freedom of DBB and MA, the active degrees of   
  	freedom can be identified. Those degrees of freedom to which most kinetic energy is imparted will be important for driving the system towards the transition state. Fig.~\ref{mdp}(a) shows that rotations contain the largest amount of kinetic energy for both DBB and MA. The rotational energy of DBB and MA together accounts for 63~\% of the total kinetic energy while translational energy accounts for 18~\%. Certain vibrations are also activated (see Fig.~S4 of the supplementary information for the individual contributions). An identical result has been obtained from the direct decomposition of the total kinetic energy (Fig.~S5 of the supplementary information). Using the sudden vector projection method \cite{guo14a} we also arrive at the conclusion that rotations are the most important degrees of freedom. This is surprising since one may naively assume that the reaction coordinate is mainly a translation and not a rotation. 	
  		
	In order to further validate this result, the MDP was also calculated using the PhysNet PES. The results are shown in Fig.~\ref{mdp}(b) and Fig.~S6 of the supporting information and qualitatively agree with those of MS-ARMD (68~\% of the total kinetic energy is imparted into rotations and 20~\% into translations), although less kinetic energy is acquired by the vibrations with PhysNet: vibrational energy accounts for 12~\% of the total kinetic energy while for MS-ARMD it is 19~\%. 
	
	The fact that the MDP between TS and reactant for MS-ARMD extends over 100~fs while for the PhysNet PES it spans for 175~fs is due to differences in the shape of the two PESs near the TS (see Fig.~S7 of the supplementary information) which define the initial accelerations of the particles.
		
	In order to investigate whether rotations are important only due to the presence of the heavy bromine atoms in the system (the large mass of bromine atoms potentially enhances the torque on DBB along the MDP), the MDP of the reaction of MA with butadiene (BD) was calculated with the PhysNet model. This is possible since the dataset on which the model was trained contains the necessary information to describe the system with hydrogen atoms instead of bromine atoms. The results are shown in Fig.~\ref{mdp}(b) (individual contributions are shown in Fig.~S8 of the supplementary information). It was found that the important degrees of freedom for MA remain approximately the same for the reaction with DBB and the reaction with BD. When comparing the distribution of kinetic energy of BD with that of DBB, it can be seen that, even though rotations are still active, translations seem to be imparted the most kinetic energy. The rotational energy of BD and MA together accounts in this case for 48~\% of the total kinetic energy while translational energy accounts for 40~\%. This indicates that the heavy bromine atoms accentuate the importance of rotational excitation for driving the reaction.

  \subsection{Cross section for the formation of van der Waals complexes in the entrance channel}
	\begin{figure}[h!]
		\includegraphics[width=0.75\textwidth]{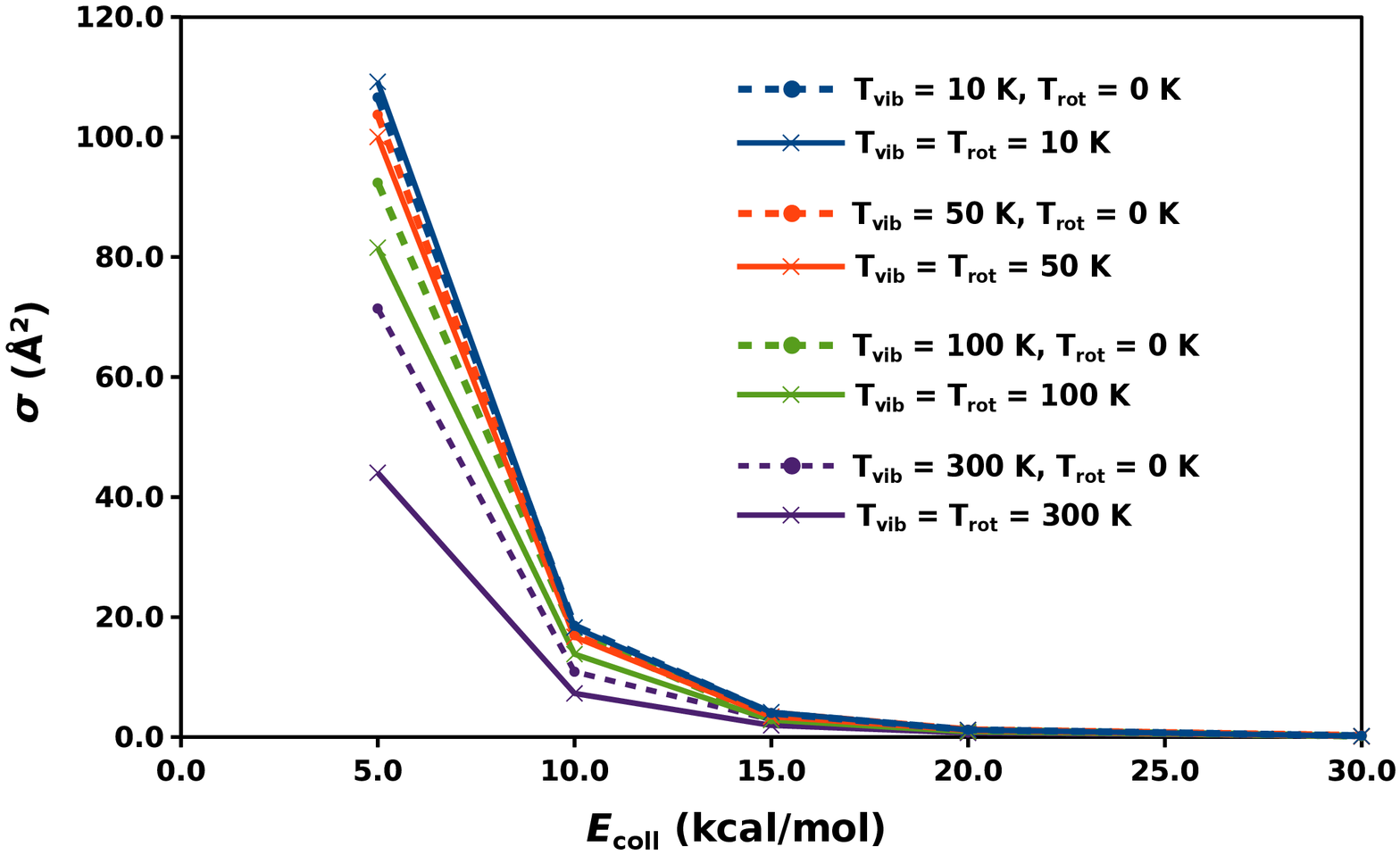} 
		\caption{Variation of the cross section ($\sigma$) for the formation of the van der Waals complex in the entrance channel of the Diels-Alder reaction between dibromobutadiene and maleic anhydride as a function of the collision energy ($E_{\rm coll}$) with different vibrational and rotational temperatures ($T_{\rm vib}$, $T_{\rm rot}$).}
		\label{vdwcross}
	\end{figure}
	The formation of van der Waals complexes in the entrance channel was studied in order to determine whether the reaction is direct (without the formation of complexes) or complex-mediated. Initial ensembles were generated as described in section~II and the impact parameter $b$ was uniformly sampled in intervals of 1~$\rm \AA$~until the maximum impact parameter was reached. For each set of initial conditions ($E_{\rm coll}$, $T_{\rm vib}$, $T_{\rm rot}$, $b$) 500~trajectories were run for 10~ps. If at the end of a trajectory, the center of mass distance between the two molecules was $<$~15~$\rm \AA$, it was considered that a van der Waals complex had been formed. Fig.~\ref{vdwcross} shows the cross section for the formation of complexes as a function of the collision energy. It can be seen that the cross section diminishes as the collision energy increases. In fact, the cross section is close to zero for $E_{\rm coll}$~$>$~20~kcal/mol. 
	
	The influence of vibrational and rotational temperature is also shown in Fig.~\ref{vdwcross}. Increasing the energy of the system in either vibrational or rotational degrees of freedom reduces the cross section for complex formation. These findings can be explained by the stabilization of the well of the van der Waals complex of around 12~kcal/mol (Fig.~\ref{neutpes}). Thus when the system has collision energies above 15~kcal/mol or sufficiently high rotational and vibrational excitation, it can dissociate or not get trapped in the potential well at all.
	
  \subsection{Reactive collisions}

	\begin{figure}[h!]
		\includegraphics[width=0.95\textwidth]{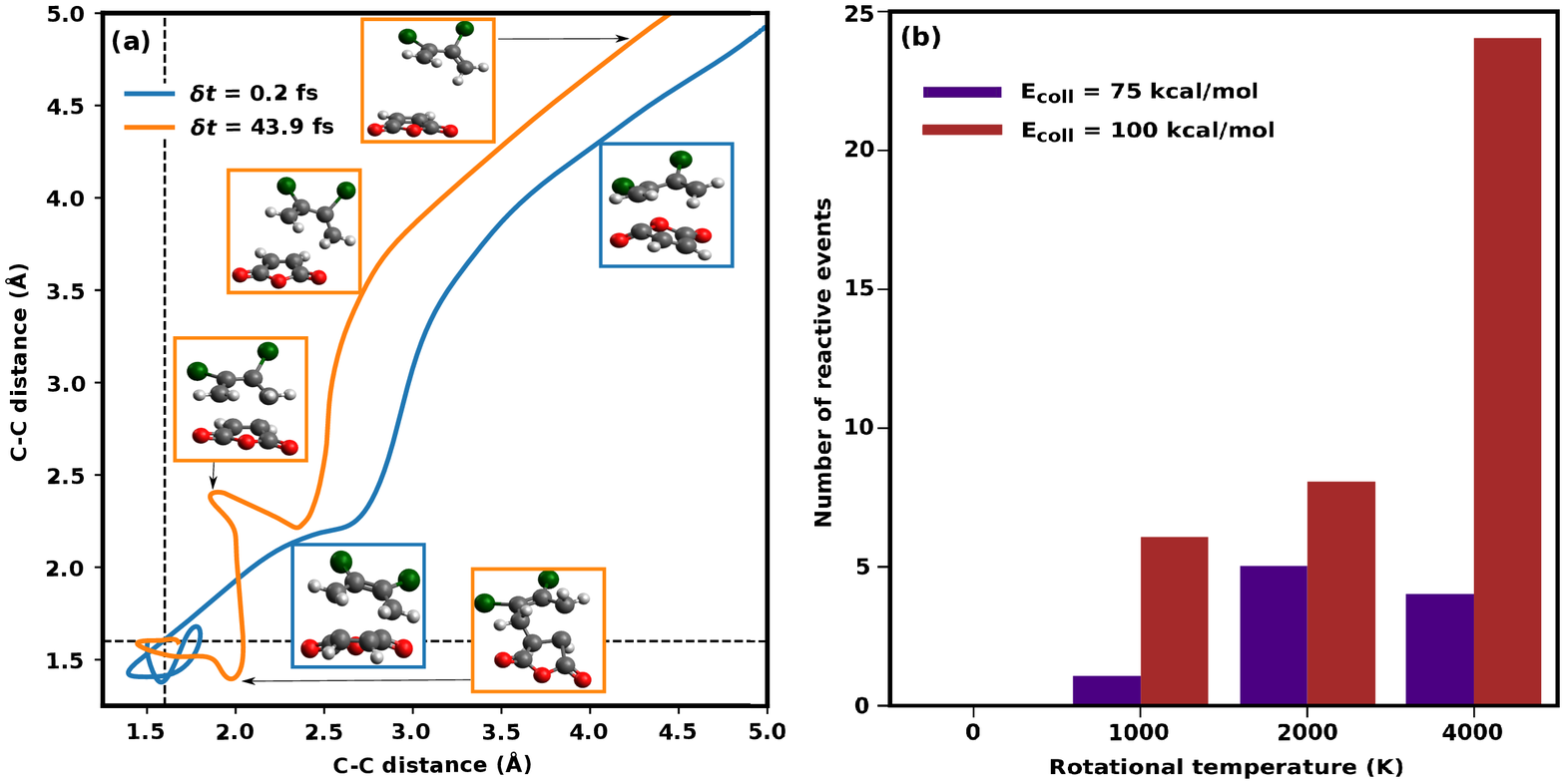}  
		\caption{(a) Distances of the two C-C bonds formed along two reactive trajectories with different times $\delta t$ elapsed between the formation of the first and the second bond. Some snapshots of structures along the trajectory are shown. The dashed, black lines at 1.6~$\rm \AA$~ indicate the geometrical threshold for a C-C bond formation. (b)~Variation of the number of reactive events at collision energies 75 and 100~kcal/mol with vibrational temperature 100~K and impact parameter 0~$\rm \AA$~as a function of the rotational temperature of the reactant molecules. 10$^5$ trajectories were run per collision energy and rotational temperature (see Table~SVIII of the supplementary information).}
		\label{reacbars}
	\end{figure}

 	For studying the Diels-Alder reaction itself, approximately $10^7$ reactive molecular dynamics simulations were carried out. The sets of simulations that yield reactive events are summarized in Table~SVIII of the supplementary information and Fig.~\ref{reacbars}(a) displays trajectories of two reactive events. The vibrational temperature was 100~K for most of the trajectories although some tests at higher vibrational temperatures have also been performed. Collision energies between 15 and 100~kcal/mol were sampled. The impact parameter was varied from 0 to 6~$\rm \AA$. The rotational temperature was 0, 1000, 2000 and 4000~K. 
 	
    In order to calculate a reactive cross section, we ran $5 \cdot 10^6$ trajectories for initial conditions $E_{\rm coll} $~=~100~kcal/mol, $T_{\rm rot}$~=~4000~K and $T_{\rm vib}$~=~100~K at which the largest number of reactive events were observed. The maximum impact parameter $b_{\rm max}$ was 5~$\rm \AA$ which yields a cross section $\sigma=2.13\cdot 10^{-3}~ \rm \AA^2$, corresponding to a rate $k=7.53\cdot 10^{-14}$~cm$^3$~molecule$^{-1}$~s$^{-1}$. A collision energy of 100~kcal/mol corresponds to a relative velocity of 3533~m/s which is at the very upper limit of what may be experimentally achievable.\cite{rivero17a}
     	
 	As the MDP shows and Fig.~\ref{reacbars}(b) confirms, the rotational energy promotes the reaction, although reactive events are still rare (1 in 10$^4$). In order to further test this result, a rotational temperature of $T_{\rm rot}$~=~8000~K was used for collisions at $E_{\rm coll}$~=~15, 20~kcal/mol but few reactive events (1 - 2) were observed, indicating that collision energy is also needed for the reaction to take place. It should be noted that only 5 reactive events out of 482 are recorded without rotational excitation (see Table~SVIII of the supplementary information). 
 	
 	With a vibrational temperature of 100~K, no reactive events could be observed with $E_{\rm coll}$~$<$~75~kcal/mol even though, in principle, with a collision energy of 10~kcal/mol the system would have enough energy to overcome the reaction barrier (Fig.~\ref{neutpes}). This might indicate that the reaction rate with such low $E_{\rm coll}$ is too small combined with the fact that at these energies the reaction could be partly complex mediated and take much longer times than simulated. For a reactive event to take place, the molecules need to collide in a suitable relative orientation in order to overcome steric constraints and with the right distribution of energy, such that the TS can be reached. From Fig.~\ref{vdwcross}, we know that there is no complex formation for $E_{\rm coll}$~$>$~20~kcal/mol and thus the reactions at the energies in Fig.~\ref{reacbars} must be direct.
 	
 	The influence of vibrational energy was tested by raising the vibrational temperature to $T_{\rm vib}$ = 1000~K and by individualy exciting some of the vibrational normal modes that appear to be active in the MDP. However, reactive events were so rare (0 - 2) that it can be concluded that vibrational activation of the reaction is weak, at least in this energy range. 
 	 
	The time $\delta t$ elapsed between formation of the first and second bond was calculated for all the reactive events in order to determine the synchronicity of the reaction (Fig.~S9 of the supplementary information). Out of 482 reactive events (see Table~SVIII of the supplementary information), only two are slightly asynchronous with time lapses of 40.4~fs and 43.9~fs which are larger than the \textit{ca.}~30~fs corresponding to a C-C bond vibrational period in cyclohexene. \cite{souza16a, diau99a} The difference between a synchronous and a slightly asynchronous process can be seen in Fig.~\ref{reacbars}(a). We have not found a correlation between the (a)symmetry of the TS structure in the dynamics and the (a)synchronicity of the process (Fig.~S9 of the supplementary information) meaning that the elapsed time does not seem to depend on the (a)symmetry of the TS structure. Here, the TS geometry along the reactive trajectory is defined as the structure at the maximum in potential energy before the systems crosses from the reactant surface to the product surface. When recrossings occur, the last crossing point is taken for the definition of the TS structure.

\section{Conclusions}

We have studied the Diels-Alder reaction between MA and DBB using reactive molecular dynamics. The trajectories start with the two reactant molecules approaching each other as in a collision experiment. The minimum dynamic path indicates that rotations are important to drive the system towards the transition state. Furthermore, this finding has been confirmed by the fact that the majority of reactive collisions occur with rotational excitation. The presence of bromine substituents in the system accentuates the importance of rotations, but they were also found to be important for reactions of non substituted dienes. At the energies at which reactive events were observed, the cross section for the formation of van der Waals complexes in the entrance channel is almost zero and thus the reaction cannot be complex-mediated. Most of the observed reactive events are synchronous.

One of the fundamental aspects in reaction dynamics concerns the question which form of energy (translation, vibration or rotation) is most efficacious for the system to reach and surmount the transition state.\cite{polanyi72a} Back in the 1970s, when studying model atom plus diatom reactions, Polanyi formulated rules which relate the nature of the transition state (early or late) with the type of energy that promotes the reaction (translational or vibrational energy). Application\cite{bowman12a} and generalization\cite{guo14a, jiang14a} of these rules to polyatomic molecules remains a challenging undertaking, see for example the sudden vector projection (SVP) model.\cite{guo14a, jiang14a} Analysis of a number of atom plus diatom (H+H$_2$, F+H$_2$, F+HCl) or atom plus triatom (H+H$_2$O, F+H$_2$O) reactions highlighted the cases under which rotations may play an important role in promoting or inhibiting a reaction.\cite{jiang14a} The strength of the SVP model is that it requires only information about the normal modes and their directions in the reactant and transition state structures. On the other hand, the ``sudden approximation'' will not be applicable to situations in which internal vibrational relaxation (IVR) occurs or for collision energies much higher than the reaction threshold, as is the case in the present work. The present work suggests that rotations can play an important role for reactions involving large excess of translational energy and the implications for reaction dynamics involving polyatomic molecules are exciting.\cite{li14c, song16a, kilaj18a, jiang14a, li14a}

As a next step, we intend to perform reactive molecular dynamics simulations for the cationic reaction between DBB and MA that is expected to be faster and in which concerted and stepwise mechanisms are anticipated to coexist.\cite{rivero17a} 

To the best of our knowledge, no simulation study had been performed in Diels-Alder reactions starting from the beginning of the reaction without steered dynamics. The present results indicate that the need of high collision energies together with rotational excitation and the low reaction rate of reaction will make the study of this reaction in single-collision experiments challenging.

\section*{Acknowledgements}

\noindent We acknowledge funding from the Swiss National Science Foundation, grant nr. BSCGI0\_157874, and the University of Basel.

\bibliography{refs_orig_o}

\end{document}


\begin{center}
\vskip-50pt
\Large{\textbf{Supplementary Information for ``Reactive Atomistic Simulations of Diels-Alder Reactions: the Importance of Molecular Rotations for Diels-Alder Reactions''}}

\vskip20pt
\large{Ux\'ia Rivero, Oliver T. Unke, Markus Meuwly, Stefan Willitsch}
\end{center}
\tableofcontents
\pagebreak
\section{Validation of the MS-ARMD model}
	\begin{figure}[H]
		\includegraphics[width=.75\textwidth]{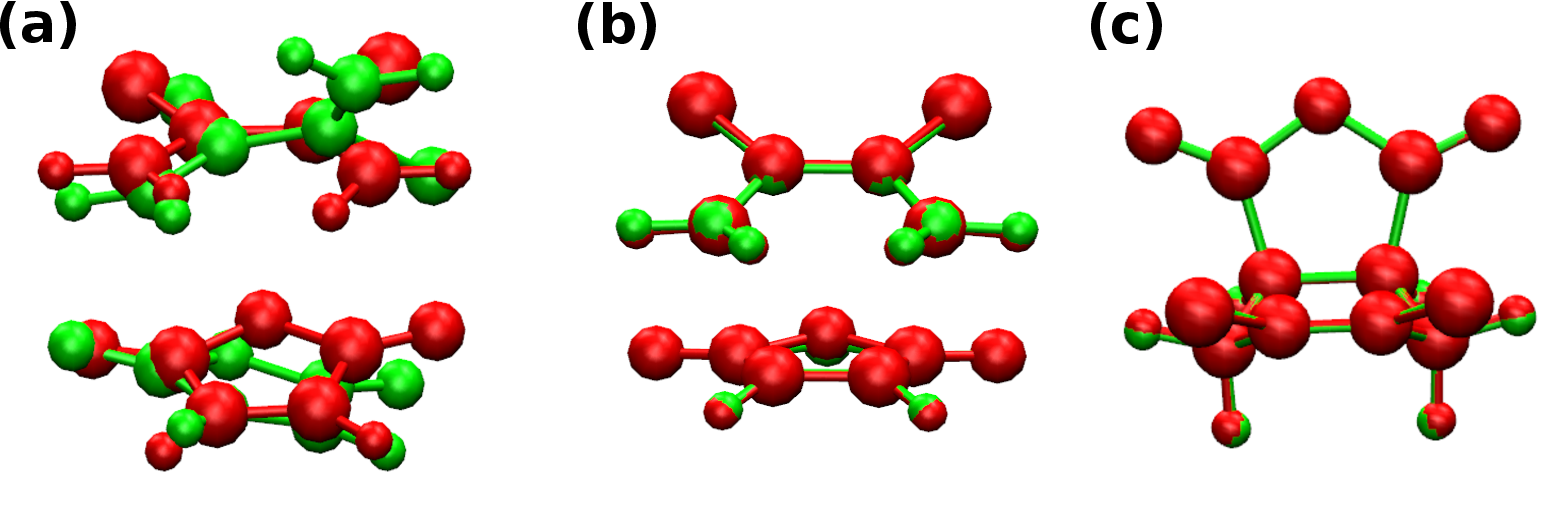}
		\caption{Comparison of (a) reactant, (b) transition state and (c) product minimized structures of the endo path at DFT (red) and MS-ARMD (green) levels of theory.} 
		\label{minstructures}
	\end{figure}

Fig.~\ref{minstructures} shows a comparison of relevant stationary points on the PES obtained at the DFT and MS-ARMD levels of theory. The main difference between the minimized structures is the C=C$-$C=C dihedral angle of DBB in the van der Waals complex (Fig.~\ref{minstructures}(a)). Describing a weakly bound complex with point charges and isotropic van der Waals interactions is an approximation. The structure of the van der Waals complex can be further improved by using multipolar interactions for the electrostatics and jointly fitting van der Waals and dihedral parameters. However, this was not further pursued because the correlation between reference and fitted energies is sufficiently good and structural details are less relevant for this work.

 	\begin{figure}[H]
 		\includegraphics[width=\textwidth]{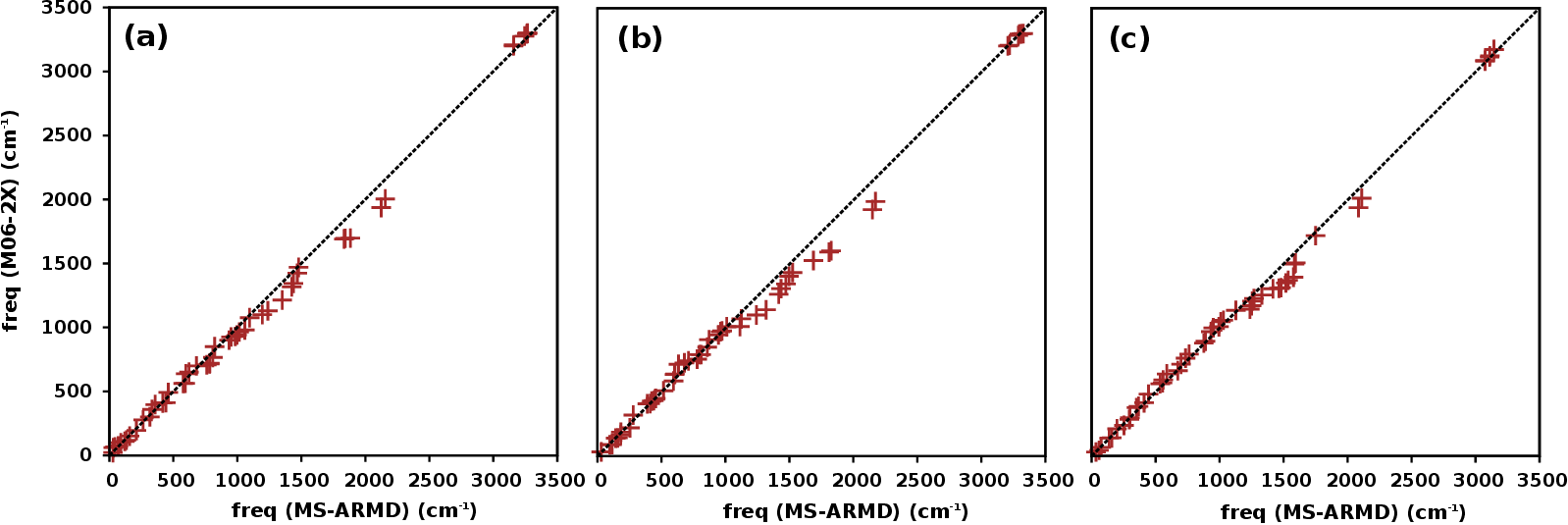}
 		\caption{Comparison of (a) reactant, (b) transition state and (c) product harmonic frequencies for the minimized structures in Fig. S1 at DFT and MS-ARMD levels of theory with RMSDs of 69.4, 85.0 and 68.5~cm$^{-1}$ respectively.} 
 		\label{frequencies}
 	\end{figure}

\pagebreak
\section{MS-ARMD model}
 	\begin{figure}[H]
 		\includegraphics[width=0.25\textwidth]{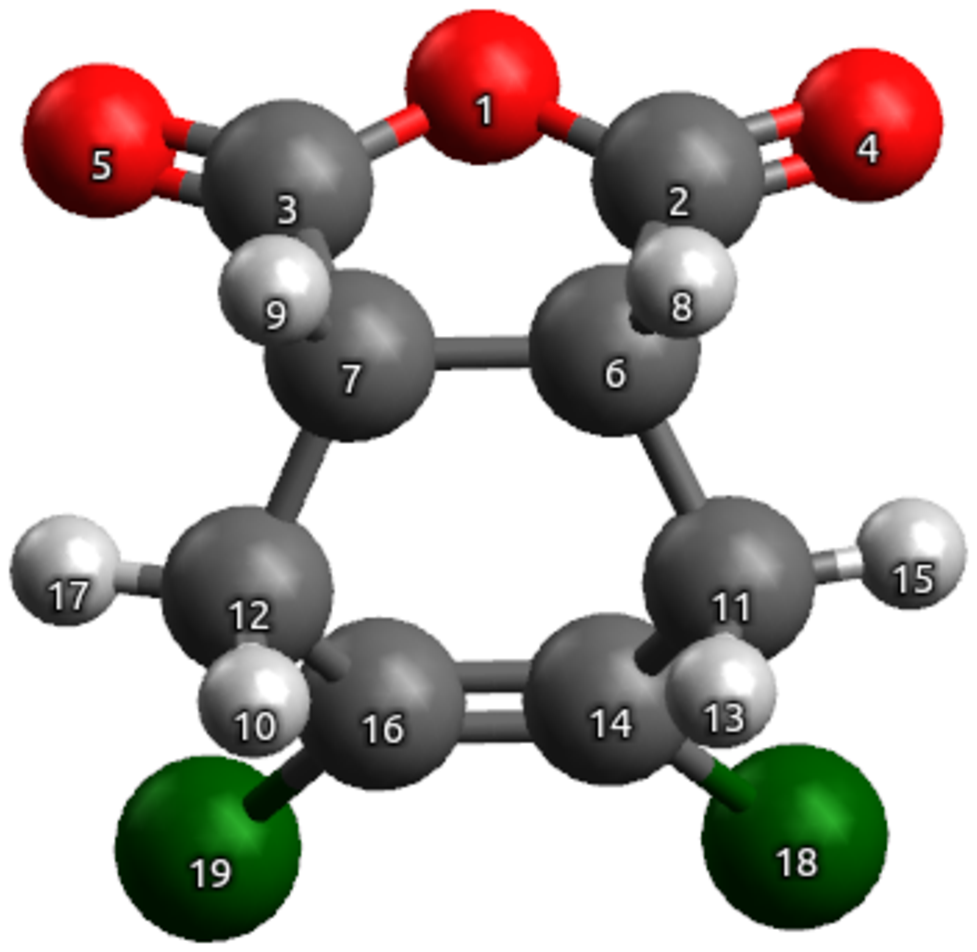}
 		\caption{Numbering of the atoms for the force fields for one of the two products force field. The other force field has the dibromobutadiene rotated such that the C-C bonds are between atoms 6-12 and 7-11. Carbons are in gray, hydrogens in white, oxygens in red and bromines in green. } 
 		\label{numbers}
 	\end{figure}


  \begin{table}[H]
    \caption{Harmonic bond parameters of the MS-ARMD reactant (columns 3 and 4)  and product (5 and 6) force fields.}
    \label{ffi}
  \begin{tabular}{c|c|c|c|c|c}

 Atom 1 \# & Atom 2 \# & $k$/2 [kcal/mol/\AA$^2$] & $r_e$ [\AA]& $k$/2 [kcal/mol/\AA$^2$] & $r_e$ [\AA]\\ \hline
2 & 4     &    1093.65 &  1.19577   &        1310.44 &  1.19499 \\
3 & 5     &    1093.65 &  1.19577   &        1310.44 &  1.19499 \\
6 & 8     &    411.539 &  1.08513   &        464.889 &  1.09667 \\
7 & 9     &    411.539 &  1.08513   &        464.889 &  1.09667 \\
11&  15   &    403.908 &  1.08680   &        464.889 &  1.09667 \\
12&  17   &    403.908 &  1.08680   &        464.889 &  1.09667 \\
11&  13   &    403.908 &  1.08680   &        464.889 &  1.09667 \\
10&  12   &    403.908 &  1.08680   &        464.889 &  1.09667 
  \end{tabular} 
  \end{table}

  \begin{table}[H]
    \caption{Morse bond parameters of the MS-ARMD reactant (columns 3, 4 and 5)  and product (6, 7 and 8) force fields. ``X'' indicates that this parameter is not needed.}
  \begin{tabular}{c|c|c|c|c|c|c|c}

 Atom 1 \# & Atom 2 \# & $D_e$ [kcal/mol] & $r_e$ [\AA]& $\beta$ [1/\AA] & $D_e$ [kcal/mol] & $r_e$ [\AA]& $\beta$ [1/\AA]\\ \hline
1 & 2    &   85.1832  & 1.39166 & 1.99423    &  62.7150 & 1.39718 & 1.98961 \\
1 & 3    &   85.1832  & 1.39166 & 1.99423    &  62.7150 & 1.39718 & 1.98961 \\
2 & 6    &   164.953  & 1.50368 & 1.40025    &  72.0637 & 1.52370 & 1.98423 \\
3 & 7    &   164.953  & 1.50368 & 1.40025    &  72.0637 & 1.52370 & 1.98423 \\
6 & 7    &   194.950  & 1.33603 & 1.99291    &  183.946 & 1.50919 & 1.07343 \\
14&  16  &   205.845  & 1.45846 & 1.40818    &  360.962 & 1.34254 & 1.48736 \\
11&  14  &   353.111  & 1.32505 & 1.47075    &  95.4170 & 1.50611 & 1.91729 \\
12&  16  &   353.111  & 1.32505 & 1.47075    &  95.4170 & 1.50611 & 1.91729 \\
7 & 12   &     X      &    X    &    X       &  183.946 & 1.50919 & 1.07343 \\
6 & 11   &     X      &    X    &    X       &  183.946 & 1.50919 & 1.07343 \\ 
14&  18  &   65.3971  & 1.89921 & 1.87225    &  68.8197 & 1.88376 & 1.96818 \\
16&  19  &   65.3971  & 1.89921 & 1.87225    &  68.8197 & 1.88376 & 1.96818 \\
                     
  \end{tabular} 
  \end{table}

  \begin{table}[H]
  \begin{tabular}{c|c|c|c|c|c|c}
  Atom 1 \# & Atom 2 \# & Atom 3 \# & $k$/2 [kcal/mol/\AA$^2$] & $\theta_e$ [\AA]& $k$/2 [kcal/mol/\AA$^2$] & $\theta_e$ [\AA]\\ \hline
2  & 1  & 3     &  64.5319 &  105.978 &   153.226 & 113.237  \\ 
1  & 2  & 4     &  99.6847 &  128.737 &   109.650 & 131.035  \\ 
1  & 2  & 6     &  135.436 &  109.951 &   99.6751 & 119.974  \\ 
4  & 2  & 6     &  77.1982 &  137.451 &   78.9705 & 140.440  \\ 
1  & 3  & 5     &  99.6847 &  128.737 &   109.650 & 131.035  \\ 
1  & 3  & 7     &  135.436 &  109.951 &   99.6751 & 119.974  \\ 
5  & 3  & 7     &  77.1982 &  137.451 &   78.9705 & 140.440  \\ 
2  & 6  & 7     &  95.5494 &  108.805 &   64.0062 & 103.976  \\ 
2  & 6  & 8     &  38.5239 &  124.711 &   50.4823 & 105.428  \\ 
2  & 6  & 11    &     X    &    X     &   64.0062 & 103.976  \\  
7  & 6  & 8     &  20.1303 &  129.751 &   56.7929 & 108.108  \\ 
7  & 6  & 11    &     X    &    X     &   77.4704 & 100.923  \\   
8  & 6  & 11    &     X    &    X     &   56.7929 & 108.108  \\  
3  & 7  & 6     &  95.5494 &  108.805 &   64.0062 & 103.976  \\ 
3  & 7  & 9     &  38.5239 &  124.711 &   50.4823 & 105.428  \\ 
3  & 7  & 12    &     X    &    X     &   64.0062 & 103.976  \\  
6  & 7  & 9     &  20.1303 &  129.751 &   56.7929 & 108.108  \\ 
6  & 7  & 12    &     X    &    X     &   77.4704 & 100.923  \\   
9  & 7  & 12    &     X    &    X     &   56.7929 & 108.108  \\   
6  & 11 &  13   &     X    &    X     &   56.7929 & 108.108  \\   
6  & 11 &  14   &     X    &    X     &   43.0251 & 101.421  \\  
6  & 11 &  15   &     X    &    X     &   56.7929 & 108.108  \\  
13 &  11&   14  &  42.9328 &  123.393 &   55.5841 & 110.000  \\ 
13 &  11&   15  &  26.6835 &  123.557 &   47.2662 & 107.155  \\ 
14 &  11&   15  &  42.9328 &  123.393 &   55.5841 & 110.000  \\ 
7  & 12 &  10   &     X    &    X     &   56.7929 & 108.108  \\  
7  & 12 &  16   &     X    &    X     &   43.0251 & 101.421  \\  
7  & 12 &  17   &     X    &    X     &   56.7929 & 108.108  \\  
  \end{tabular} 
  \end{table} 
  
 \begin{table}[H]
 \caption{Angle parameters of the MS-ARMD reactant (columns 4 and 5)  and product (6 and 7) force fields. ``X'' indicates that this parameter is not needed. }
  \begin{tabular}{c|c|c|c|c|c|c} 
  Atom 1 \# & Atom 2 \# & Atom 3 \# & $k$/2 [kcal/mol/\AA$^2$] & $\theta_e$ [\AA]& $k$/2 [kcal/mol/\AA$^2$] & $\theta_e$ [\AA]\\ \hline
10 &  12&   16  &  42.9328 &  123.393 &   55.5841 & 110.000  \\ 
10 &  12&   17  &  26.6835 &  123.557 &   47.2662 & 107.155  \\ 
16 &  12&   17  &  42.9328 &  123.393 &   55.5841 & 110.000  \\ 
11 &  14&   16  &  38.6304 &  128.237 &   55.9672 & 122.309  \\ 
11 &  14&   18  &  64.6768 &  127.307 &   69.3366 & 115.166  \\ 
16 &  14&   18  &  55.6080 &  121.871 &   87.6601 & 124.293  \\ 
12 &  16&   14  &  38.6304 &  128.237 &   55.9672 & 122.309  \\ 
12 &  16&   19  &  64.6768 &  127.307 &   69.3366 & 115.166  \\ 
14 &  16&   19  &  55.6080 &  121.871 &   87.6601 & 124.293  \\                   
  \end{tabular} 
  \end{table}

   \begin{table}[H] 
   \begin{tabular}{c|c|c|c|c|c|c|c}
   Atom 1 \# & Atom 2 \# & Atom 3 \# & Atom 4 \# & N & $k$ [kcal/mol]& $k$ [kcal/mol] & $\phi$ [degree]\\ \hline
 1 &   2 &   6  &   7      &   1    &    -0.071       &      -0.456229    &     0.00     \\
 1 &   2 &   6  &   7      &   2    &    4.79788      &      1.33681      &   180.00     \\
 1 &   2 &   6  &   7      &   3    &        X        &      0.725632     &   0.00       \\
 1 &   2 &   6  &   8      &   1    &     0.179       &         X         &   0.00       \\ 
 1 &   2 &   6  &   8      &   2    &    0.249864     &      1.52218      &   180.00     \\
 1 &   2 &   6  &   8      &   3    &     0.097       &      1.52546      &  0.00        \\
 1 &   2 &   6  &   11     &   1    &        X        &     -0.456229     &    0.00      \\
 1 &   2 &   6  &   11     &   2    &        X        &      1.33681      &   180.00     \\
 1 &   2 &   6  &   11     &   3    &        X        &      0.725632     &   0.00       \\   
 1 &   3 &   7  &   6      &   1    &    -0.071       &      -0.456229    &     0.00     \\
 1 &   3 &   7  &   6      &   2    &    4.79788      &      1.33681      &   180.00     \\
 1 &   3 &   7  &   6      &   3    &        X        &      0.725632     &   0.00       \\
 1 &   3 &   7  &   9      &   1    &     0.179       &             X     &    0.00      \\  
 1 &   3 &   7  &   9      &   2    &     0.249864    &      1.52218      &   180.00     \\
 1 &   3 &   7  &   9      &   3    &     0.097       &      1.52546      &  0.00        \\
 1 &   3 &   7  &   12     &   1    &        X        &     -0.456229     &    0.00      \\
 1 &   3 &   7  &   12     &   2    &        X        &      1.33681      &   180.00     \\
 1 &   3 &   7  &   12     &   3    &        X        &      0.725632     &   0.00       \\   
 2 &   1 &   3  &   5      &   1    &    0.701169     &      2.91529      &  0.00        \\
 2 &   1 &   3  &   5      &   2    &    6.21522      &      1.47165      &  180.00      \\
 2 &   1 &   3  &   5      &   3    &    0.298850     &      0.909140     &    0.00      \\
 2 &   1 &   3  &   7      &   2    &    7.17573      &      4.31208      &  180.00      \\
 2 &   6 &   7  &   3      &   1    &        X        &     -1.41346      &  0.00        \\
 2 &   6 &   7  &   3      &   2    &    8.19928      &          X        &  180.00      \\     
 2 &   6 &   7  &   3      &   3    &        X        &      -2.99571     &    0.00      \\
 2 &   6 &   7  &   9      &   1    &        X        &       1.82442     &    0.00      \\
 2 &   6 &   7  &   9      &   2    &    6.34212      &      -2.00004     &   180.00     \\
 2 &   6 &   7  &   12     &   1    &        X        &      -1.95087     &   0.00       \\

   \end{tabular} 
   \end{table} 
   \begin{table}[H] 
   \begin{tabular}{c|c|c|c|c|c|c|c}
   Atom 1 \# & Atom 2 \# & Atom 3 \# & Atom 4 \# & N & $k$ [kcal/mol]& $k$ [kcal/mol] & $\phi$ [degree]\\ \hline   
 2 &   6 &   7  &   12     &   2    &        X        &      -0.959448    &     180.00   \\
 2 &   6 &   7  &   12     &   3    &        X        &       0.307973    &    0.00      \\  
 2 &   6 &   11 &    13    &   1    &        X        &       1.82442     &    0.00      \\
 2 &   6 &   11 &    13    &   2    &        X        &      -2.00004     &   180.00     \\ 
 2 &   6 &   11 &    14    &   3    &        X        &      -2.07692     &   0.00       \\ 
 2 &   6 &   11 &    15    &   1    &        X        &       1.82442     &    0.00      \\
 2 &   6 &   11 &    15    &   2    &        X        &      -2.00004     &   180.00     \\ 
3  &  1  &  2   &  4       &  1     &   0.701169      &     2.91529       & 0.00         \\ 
3  &  1  &  2   &  4       &  2     &   6.21522       &     1.47165       & 180.00       \\ 
3  &  1  &  2   &  4       &  3     &   0.298850      &     0.909140      &   0.00       \\ 
3  &  1  &  2   &  6       &  2     &   7.17573       &     0.909140      &  180.00      \\ 
3  &  7  &  6   &  8       &  1     &       X         &      1.82442      &   0.00       \\ 
3  &  7  &  6   &  8       &  2     &   6.34212       &     -2.00004      &  180.00      \\ 
3  &  7  &  6   &  11      &  1     &       X         &     -1.95087      &  0.00        \\ 
3  &  7  &  6   &  11      &  2     &       X         &     -0.959448     &    180.00    \\ 
3  &  7  &  6   &  11      &  3     &       X         &     0.307973      &  0.00        \\      
3  &  7  &  12  &   10     &  1     &       X         &     1.82442       &  0.00        \\ 
3  &  7  &  12  &   10     &  2     &       X         &     -2.00004      &  180.00      \\  
3  &  7  &  12  &   16     &  3     &       X         &     -2.07692      &  0.00        \\   
3  &  7  &  12  &   17     &  1     &       X         &     1.82442       &  0.00        \\ 
3  &  7  &  12  &   17     &  2     &       X         &     -2.00004      &  180.00      \\  
4  &  2  &  6   &  7       &  1     &    0.181        &     -0.224498E-01 &       0.00   \\ 
4  &  2  &  6   &  7       &  2     &   -0.999624     &     1.23963       & 180.00       \\ 
4  &  2  &  6   &  7       &  3     &       X         &     -0.838198E-01 &       0.00   \\ 
4  &  2  &  6   &  8       &  1     &       X         &     -2.02295      &  0.00        \\ 
4  &  2  &  6   &  8       &  2     &   2.45356       &     -0.342985     &    180.00    \\ 
4  &  2  &  6   &  8       &  3     &       X         &     -0.853726E-01 &       0.00   \\  
4  &  2  &  6   &  11      &  1     &       X         &     -0.224498E-01 &       0.00   \\  
   \end{tabular} 
   \end{table} 
   \begin{table}[H] 
   \begin{tabular}{c|c|c|c|c|c|c|c}
   Atom 1 \# & Atom 2 \# & Atom 3 \# & Atom 4 \# & N & $k$ [kcal/mol]& $k$ [kcal/mol] & $\phi$ [degree]\\ \hline  

4  &  2  &  6   & 11      &  2     &       X          &       1.23963       & 180.00       \\  
4  &  2  &  6   & 11      &  3     &       X          &       -0.838198E-01 &       0.00   \\   
5  &  3  &  7   & 6       &  1     &    0.181         &       -0.224498E-01 &       0.00   \\  
   5  &  3  &  7   & 6       &  2     &   -0.999624      &       1.23963       & 180.00       \\  
   5  &  3  &  7   & 6       &  3     &       X          &       -0.838198E-01 &       0.00   \\  
5  &  3  &  7   & 9       &  1     &       X          &       -2.02295      &  0.00        \\  
5  &  3  &  7   & 9       &  2     &   2.45356        &       -0.342985     &    180.00    \\  
5  &  3  &  7   & 9       &  3     &       X          &       -0.853726E-01 &       0.00   \\  
5  &  3  &  7   & 12      &  1     &       X          &       -0.224498E-01 &       0.00   \\  
5  &  3  &  7   & 12      &  2     &       X          &       1.23963       & 180.00       \\  
5  &  3  &  7   & 12      &  3     &       X          &       -0.838198E-01 &       0.00   \\   
6  &  7  &  12  &  10     &  1     &       X          &       -1.73754      &  0.00        \\  
6  &  7  &  12  &  10     &  2     &       X          &       -0.739873     &    180.00    \\  
6  &  7  &  12  &  10     &  3     &       X          &       0.842630E-01  &      0.00    \\   
6  &  7  &  12  &  16     &  1     &       X          &       -0.800732     &    0.00      \\
6  &  7  &  12  &  16     &  2     &       X          &       -1.97784      &  180.00      \\
6  &  7  &  12  &  16     &  3     &       X          &       0.824677      &  0.00        \\ 
6  &  7  &  12  &  17     &  1     &       X          &       -1.73754      &  0.00        \\
6  &  7  &  12  &  17     &  2     &       X          &       -0.739873     &    180.00    \\
6  &  7  &  12  &  17     &  3     &      X           &       0.842630E-01  &      0.00    \\ 
6  &  11 &   14 &   16    &  1     &       X          &       -0.758787     &    0.00      \\
6  &  11 &   14 &   16    &  2     &       X          &       -0.416878     &   180.00     \\
6  &  11 &   14 &   16    &  3     &       X          &       -0.739873     &    0.00      \\ 
6  &  11 &   14 &   18    &  1     &       X          &       0.000         &  0.00        \\     
7  &  6  &  11  &  13     &  1     &       X          &       -1.73754      &  0.00        \\
7  &  6  &  11  &  13     &  2     &       X          &       -0.739873     &    180.00    \\
7  &  6  &  11  &  13     &  3     &       X          &       0.842630E-01  &      0.00    \\ 
7  &  6  &  11  &  14     &  1     &       X          &       -0.800732     &    0.00      \\

   \end{tabular} 
   \end{table} 
   \begin{table}[H] 
   \begin{tabular}{c|c|c|c|c|c|c|c}
   Atom 1 \# & Atom 2 \# & Atom 3 \# & Atom 4 \# & N & $k$ [kcal/mol]& $k$ [kcal/mol] & $\phi$ [degree]\\ \hline  

7  &  6  &  11  &  14     &  2     &       X          &       -1.97784      &  180.00      \\
7  &  6  &  11  &  14     &  3     &       X          &       0.824677      &  0.00        \\ 
7  &  6  &  11  &  15     &  1     &       X          &       -1.73754      &  0.00        \\
7  &  6  &  11  &  15     &  2     &       X          &       -0.739873     &    180.00    \\
7  &  6  &  11  &  15     &  3     &       X          &       0.842630E-01  &      0.00    \\ 
7  &  12 &   16 &   14    &  1     &       X          &       -0.800732     &    0.00      \\
7  &  12 &   16 &   14    &  2     &       X          &       -0.416878     &   180.00     \\
7  &  12 &   16 &   14    &  3     &       X          &       -0.739873     &    0.00      \\ 
7  &  12 &   16 &   19    &  1     &       X          &       0.000         &   0.00       \\      
8  &  6  &  7   & 9       &  1     &       X          &       0.343877      &  0.00        \\  
8  &  6  &  7   & 9       &  2     &   2.34735        &       -1.99591      &   180.00     \\  
8  &  6  &  7   & 9       &  3     &       X          &       -0.136668     &   0.00       \\  
8  &  6  &  7   & 12      &  1     &       X          &       -1.73754      &  0.00        \\  
8  &  6  &  7   & 12      &  2     &       X          &       -0.739873     &    180.00    \\  
8  &  6  &  7   & 12      &  3     &       X          &       0.842630E-01  &      0.00    \\     
8  &  6  &  11  &  13     &  1     &       X          &       0.343877      &  0.00        \\  
8  &  6  &  11  &  13     &  2     &       X          &       -1.99591      &   180.00     \\  
8  &  6  &  11  &  13     &  3     &       X          &       -0.136668     &   0.00       \\   
8  &  6  &  11  &  14     &  1     &       X          &       1.47393       & 0.00         \\  
8  &  6  &  11  &  14     &  2     &       X          &       -0.225128     &    180.00    \\  
8  &  6  &  11  &  14     &  3     &       X          &       0.355477      &  0.00        \\     
8  &  6  &  11  &  15     &  1     &       X          &       0.343877      &  0.00        \\  
8  &  6  &  11  &  15     &  2     &       X          &       -1.99591      &   180.00     \\  
8  &  6  &  11  &  15     &  3     &       X          &       -0.136668     &   0.00       \\   
9  &  7  &  6   & 11      &  1     &       X          &       -1.73754      &  0.00        \\  
9  & 7   & 6    & 11    &    2       &     X         &           -0.739873     &   180.00     \\
9  & 7   & 6    & 11    &    3       &     X         &           0.842630E-01  &     0.00     \\
9  & 7   & 12   & 10    &   1        &    X          &         0.343877      &   0.00       \\  

   \end{tabular} 
   \end{table} 
   \begin{table}[H] 
   \begin{tabular}{c|c|c|c|c|c|c|c}
   Atom 1 \# & Atom 2 \# & Atom 3 \# & Atom 4 \# & N & $k$ [kcal/mol]& $k$ [kcal/mol] & $\phi$ [degree]\\ \hline  
9  & 7   & 12   & 10    &   2        &    X          &         -1.99591      &    180.00    \\  
9  & 7   & 12   & 10    &   3        &    X          &         -0.136668     &    0.00      \\   
9  & 7   & 12   & 16    &   1        &    X          &         1.47393       &  0.00        \\  
9  & 7   & 12   & 16    &   2        &    X          &         -0.225128     &     180.00   \\  
9  & 7   & 12   & 16    &   3        &    X          &         0.355477      &   0.00       \\   
9  & 7   & 12   & 17    &   1        &    X          &         0.343877      &   0.00       \\  
9  & 7   & 12   & 17    &   2        &    X          &         -1.99591      &    180.00    \\  
9  & 7   & 12   & 17    &   3        &    X          &         -0.136668     &    0.00      \\ 
10 &  12 &   16 &   14  &   1        &    X          &         -0.775003E-01 &        0.00  \\    
10 &  12 &   16 &   14  &       2    &    4.77285    &         -0.225128     &     180.00   \\    
10 &  12 &   16 &   14  &   3        &    X          &         -0.876922     &     0.00     \\    
10 &  12 &   16 &   19  &   1        &    X          &         0.000         &    0.00      \\         
10 &  12 &   16 &   19  &       2    &    6.16283    &           X           &    180.00    \\             
11 &  6  &  7   & 12    &   1        &    X          &         -0.314136     &    0.00      \\    
11 &  6  &  7   & 12    &   2        &    X          &         3.05348       &    180.00    \\      
11 &  6  &  7   & 12    &   3        &    X          &         -0.796867     &    0.00      \\      
11 &  14 &   16 &   12  &       1    &    -5.25896   &         -1.28302      &    0.00      \\    
11 &  14 &   16 &   12  &       2    &    -0.525596  &         6.94818       &   180.00     \\     
11 &  14 &   16 &   12  &       3    &    0.085522   &              X        &   0.00       \\       
11 &  14 &   16 &   19  &       2    &    0.143296   &         6.94818       &   180.00     \\     
12 &  16 &   14 &   18  &       2    &    0.143296   &         6.94818       &   180.00     \\     
13 &  11 &   14 &   16  &   1        &    X          &         -0.775003E-01 &     0.00     \\ 
13 &  11 &   14 &   16  &       2    &    4.77285    &         -0.225128     &     180.00   \\    
13 &  11 &   14 &   16  &   3        &    X          &         -0.876922     &     0.00     \\    
13 &  11 &   14 &   18  &   1        &    X          &         0.000         &     0.00     \\          
13 &  11 &   14 &   18  &       2    &    6.16283    &            X          &    180.00    \\            
14 &  16 &   12 &   17  &   1        &    X          &         -0.775003E-01 &        0.00  \\    
   \end{tabular} 
   \end{table} 
   \begin{table}[H] 
   \caption{Dihedral parameters of the MS-ARMD reactant (columns 6)  and product (7) force fields.``X'' indicates that this parameter is not needed. }
   \begin{tabular}{c|c|c|c|c|c|c|c}
   Atom 1 \# & Atom 2 \# & Atom 3 \# & Atom 4 \# & N & $k$ [kcal/mol]& $k$ [kcal/mol] & $\phi$ [degree]\\ \hline  
14 &  16 &   12 &   17  &       2    &    4.26042    &         -0.225128     &     180.00   \\    
14 &  16 &   12 &   17  &   3        &    X          &         -0.876922     &     0.00     \\    
15 &  11 &   14 &   16  &   1        &    X          &         -0.775003E-01 &        0.00  \\    
15 &  11 &   14 &   16  &       2    &    4.77285    &         -0.225128     &     180.00   \\    
15 &  11 &   14 &   16  &   3        &    X          &         -0.876922     &     0.00     \\    
15 &  11 &   14 &   18  &   1        &    X          &         0.000         &    0.00      \\         
15 &  11 &   14 &   18  &       2    &    6.16283    &             X         &   180.00     \\          
17 &  12 &   16 &   19  &   1        &    X          &         0.000         &   0.00       \\        
17 &  12 &   16 &   19  &       2    &    6.16283    &             X         &   180.00     \\          
18 &  14 &   16 &   19  &       2    &    0.350057   &         6.94818       &   180.00     \\     
         
   \end{tabular} 
   \end{table}

   \begin{table}[H] 
   \caption{Non bonded parameters of the MS-ARMD reactant force field. ``X'' indicates that this parameter is not needed.}
   \begin{tabular}{c|c|c|c|c|c} 
   Atom \# & $q_i$ [e] & $\epsilon_{i,1}$ [kcal/mol]& $R_{min,1}$/2[\AA] &$\epsilon_{i,2}$ [kcal/mol]& $R_{min,2}$/2[\AA] \\ \hline  
1   &     -0.300000 &  0.725240     &  1.52534  &  X      &  X          \\
2   &      0.705600 &  0.100119E-05 &  3.49075  &  X      &  X          \\
3   &      0.705600 &  0.100119E-05 &  3.49075  &  X      &  X          \\
4   &     -0.570000 &  0.116541E-01 &  1.81590  &  0.12   & 1.40        \\
5   &     -0.570000 &  0.116541E-01 &  1.81590  &  0.12   & 1.40        \\
6   &     -0.135600 &  0.180888E-01 &  2.02067  &  X      &  X          \\
7   &     -0.135600 &  0.180888E-01 &  2.02067  &  X      &  X          \\
8   &      0.150000 &  0.146019     &  1.26142  &  X      &  X          \\
9   &      0.150000 &  0.146019     &  1.26142  &  X      &  X          \\
10  &      0.150000 &  0.146019     &  1.26142  &  X      &  X          \\
11  &     -0.300000 &  0.180888E-01 &  2.02067  &  X      &  X          \\
12  &     -0.300000 &  0.180888E-01 &  2.02067  &  X      &  X          \\
13  &      0.150000 &  0.146019     &  1.26142  &  X      &  X          \\
14  &      0.110000 &  0.180888E-01 &  2.02067  &  X      &  X          \\
15  &      0.150000 &  0.146019     &  1.26142  &  X      &  X          \\
16  &      0.110000 &  0.180888E-01 &  2.02067  &  X      &  X          \\
17  &      0.150000 &  0.146019     &  1.26142  &  X      &  X          \\
18  &     -0.110000 &  6.09988      &  1.75471  &  X      &  X          \\
19  &     -0.110000 &  6.09988      &  1.75471  &  X      &  X          \\ \hline

Atom 1 \# & Atom 2\# & $\epsilon_{i}$ [kcal/mol] & $R_{min}$/2[\AA] & $n$ & $m$ \\ \hline
18 &   4 &  2.03805   &   3.29190   &   13.1468  &    16.0708 \\  
18 &   5 &  2.03805   &   3.29190   &   13.1468  &    16.0708 \\  
19 &   4 &  2.03805   &   3.29190   &   13.1468  &    16.0708 \\  
19 &   5 &  2.03805   &   3.29190   &   13.1468  &    16.0708 \\  
11 &   6 &  1.84792   &   1.87951   &   3.24440  &    5.14600 \\  
12 &   7 &  1.84792   &   1.87951   &   3.24440  &    5.14600 \\  
11 &   7 &  1.84792   &   1.87951   &   3.24440  &    5.14600 \\  
12 &   6 &  1.84792   &   1.87951   &   3.24440  &    5.14600 \\

   \end{tabular} 
   \end{table}       
      
   \begin{table}[H] 
   \caption{Non bonded parameters of the MS-ARMD products force field. ``X'' indicates that this parameter is not needed.} 
   \begin{tabular}{c|c|c|c|c|c} 
   Atom \# & $q_i$ [e] & $\epsilon_{i,1}$ [kcal/mol]& $R_{min,1}$/2[\AA] &$\epsilon_{i,2}$ [kcal/mol]& $R_{min,2}$/2[\AA] \\ \hline 
1    &   -0.300000     & 0.152100 & 1.770000 &  X    &  X   \\
2    &    0.659000     & 0.110000 & 2.000000 &  X    &  X   \\        
3    &    0.659000     & 0.110000 & 2.000000 &  X    &  X   \\        
4    &   -0.570000     & 0.120000 & 1.700000 &  0.12 & 1.40 \\ 
5    &   -0.570000     & 0.120000 & 1.700000 &  0.12 & 1.40 \\ 
6    &    0.610000E-01 & 0.055000 & 2.175000 &  0.01 & 1.90 \\ 
7    &    0.610000E-01 & 0.055000 & 2.175000 &  0.01 & 1.90 \\ 
8    &    0.00000      & 0.022000 & 1.320000 &  X    &  X   \\        
9    &    0.00000      & 0.022000 & 1.320000 &  X    &  X   \\        
10   &    0.00000      & 0.022000 & 1.320000 &  X    &  X   \\        
11   &    0.138200     & 0.055000 & 2.175000 &  0.01 & 1.90 \\ 
12   &    0.138200     & 0.055000 & 2.175000 &  0.01 & 1.90 \\ 
13   &    0.00000      & 0.022000 & 1.320000 &  X    &  X   \\        
14   &   -0.282000E-01 & 0.068000 & 2.090000 &  X    &  X   \\        
15   &    0.00000      & 0.022000 & 1.320000 &  X    &  X   \\        
16   &   -0.282000E-01 & 0.068000 & 2.090000 &  X    &  X   \\        
17   &    0.00000      & 0.022000 & 1.320000 &  X    &  X   \\        
18   &   -0.110000     & 5.48335  & 1.54929  &  X    &  X   \\        
19   &   -0.110000     & 5.48335  & 1.54929  &  X    &  X   \\     
   \end{tabular} 
   \end{table}    
   
   \newpage
\begin{flushleft}
   The barrier region connecting the reactant and product state is described by two GAPOs
   $$ \Delta V^{ij}_{GAPO,k} (x) = exp\left( -\frac{(\Delta V_{ij}(x)-V^0_{ij,k})^2}{2 \sigma^2_{ij,k}}\right) \times \sum_{l=0}^{m_{ij,k}} a_{ij,kl} (\Delta V_{ij}(x)-V^0_{ij,k})^l$$
   with the parameters summarized in Table~\ref{ffend}.
\end{flushleft}
   \begin{table}[htb] 
   \caption{GAPO parameters: $i$ labels the reactant, $j$ labels the product, $V^0_{ij,k}$  is the center of the Gaussian function (in kcal/mol), and $\sigma_{ij,k}$ the width of the Gaussian (in kcal/mol). $a_{ij}$ is the polynomial coefficient in kcal/mol $^{1 - j}$, j = 0, 1.}
   \label{ffend}
   \begin{tabular}{c|c|c|c|c} 
    $k$ & $V^0_{ij,k}$ & $\sigma_{ij,k}$ & $a_{ij,k0}$ & $a_{ij,k1}$ \\ \hline
     1 & -3.2889475989E+01 & 7.7839312146E+01 & -4.0222992796E+01 &-2.4658081680E-01 \\
     0 &  1.4151964786E+00 & 2.6626692901E+01 & -9.5075322684E+00 & {} \\
   
   \end{tabular} 
   \end{table}

\section{Decomposition of the total kinetic energy along the MDP}
	\begin{figure}[H]
		\includegraphics[width=.5\textwidth]{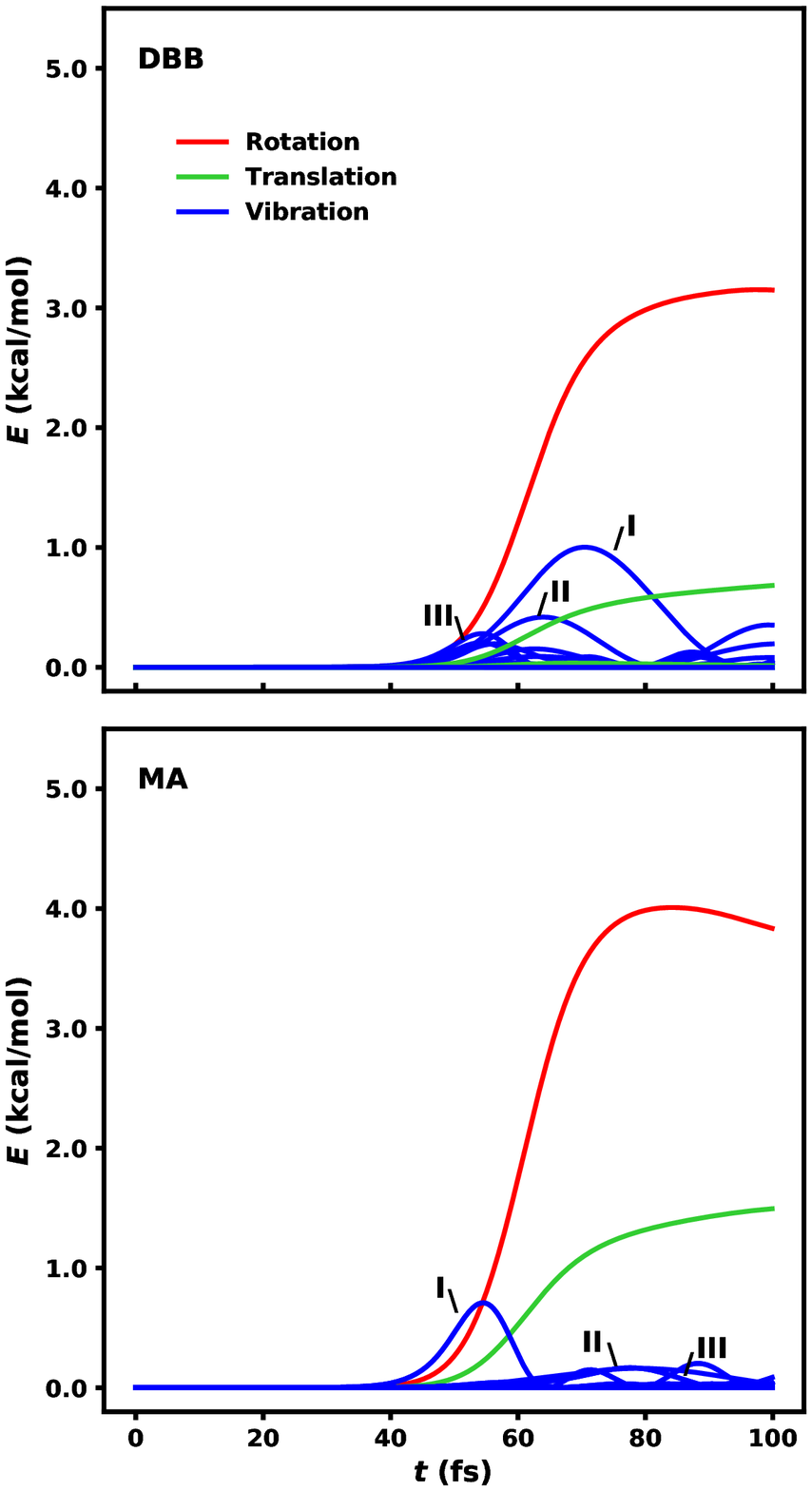}
		\caption{Projection of the total kinetic energy ($E$) along the minimum dynamic path calculated with the MS-ARMD model onto the degrees of freedom of dibromobutadiene (DBB) and maleic anhydride (MA). The trajectory starts at the endo transition state and ends at the reactants. The most important vibrations for DBB are (I) molecular out-of-plane bending and (II, III) C-H out-of-plane bending mixed with skeleton vibrations. For MA they are (I) C-C symmetric stretching, (II) C=O out-of-plane symmetric bending and (III) out-of-plane bending of the molecule.} 
		\label{msarmd-all}
	\end{figure} 

\newpage
The decomposition of the total kinetic energy into rotational, translational and vibrational energy of MA and DBB has also been calculated directly from the atomic velocities. The translational energy ($E_{\rm trans}$) is the kinetic energy of the centers of mass of MA and DBB (Eq.~\ref{etrans}) where $\vec{p}_i$ is the momentum of atom $i$ belonging to molecule $j$ ($j$ = MA, DBB) and $M_j$ is the total mass of molecule $j$. 

\begin{equation}
  E_{{\rm trans}, j} =\frac{|\sum_{i \in j} \vec{p}_i|^2}{2 \cdot M_j}
 \label{etrans}
\end{equation}
The rotational energy ($E_{\rm rot}$) is calculated as stated in Equations~\ref{erot1}-\ref{erot2} where $\vec{x}$ can be atomic momenta ($\vec{p}$), atomic coordinates ($\vec{r}$) or atomic velocities ($\vec{v}$), the subscript ${\rm CoM},j$ refers to the center of mass of molecule $j$, $\vec{L}_j$ is the angular momentum of molecule $j$, $\vec{\omega}_j$ is the angular velocity of molecule $j$ and $\textbf{I}_j$ is the inertia tensor of molecule $j$.
\begin{equation}
  {\vec{x}_i}~' = \vec{x}_i - \vec{x}_{{\rm CoM}, j}
  \label{erot1}
\end{equation}

\begin{equation}
  \vec{L}_j = \sum_{i \in j} \vec{r}_i~' \times \vec{p}_i~'
  \label{erot3}
\end{equation}

\begin{equation}
  \vec{\omega}_j = \textbf{I}_j^{-1} \vec{L}_j
  \label{erot4}
\end{equation}

\begin{equation}
  E_{{\rm rot}, j} =\frac{1}{2} | \textbf{I}_j \vec{\omega}_j^2|
 \label{erot2}
\end{equation}

The vibrational kinetic energy ($E_{\rm vib}$) is calculated from the total kinetic energy of the individual molecules ($E_{\rm tot}$) as in Eq.~\ref{evib}
\begin{equation}
   E_{{\rm vib}, j} = E_{{\rm tot}, j} - E_{{\rm rot}, j} - E_{{\rm trans}, j}
  \label{evib}
\end{equation}

	\begin{figure}[H]
		\includegraphics[width=.5\textwidth]{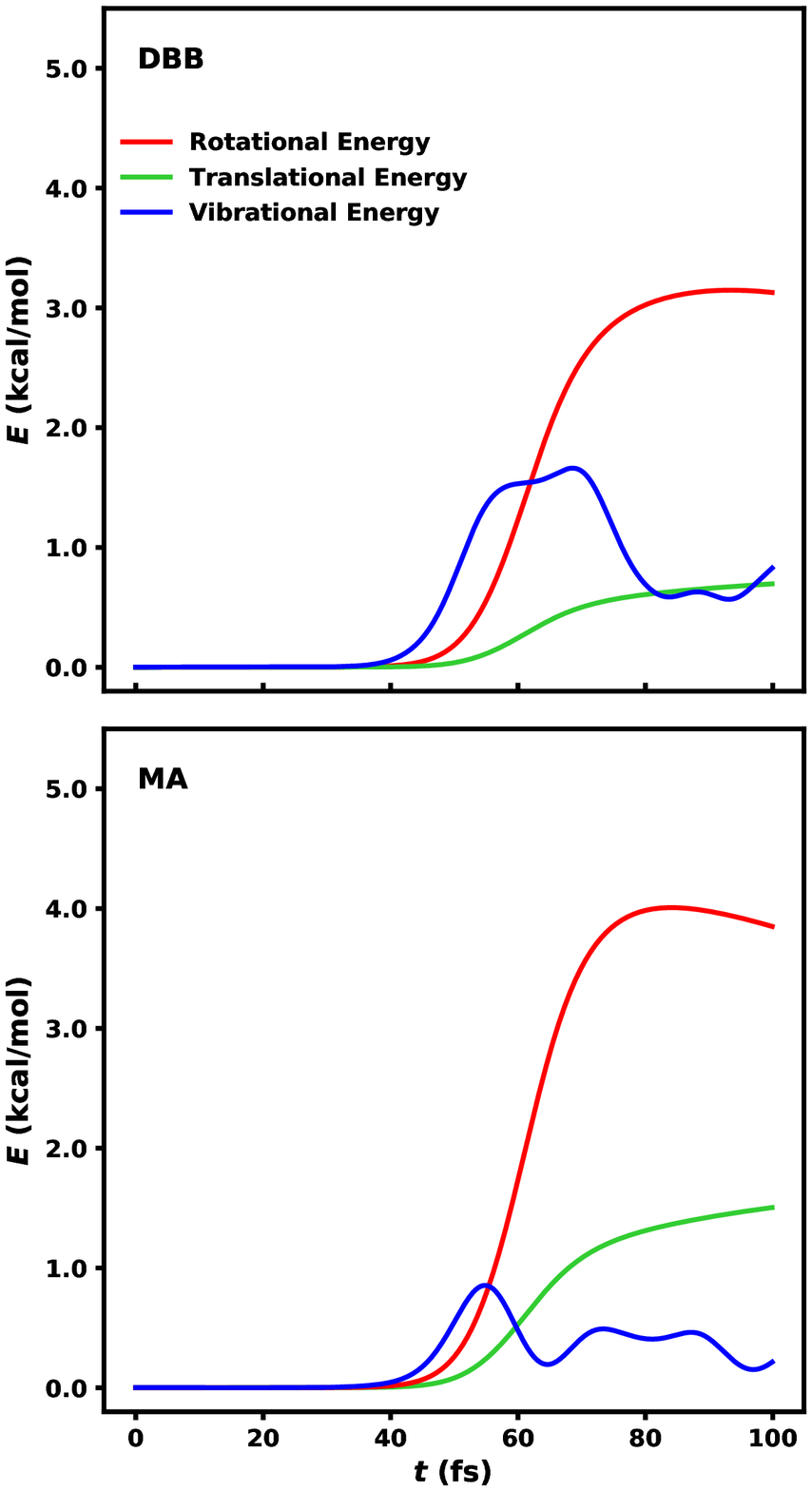}
		\caption{Decomposition of the total kinetic energy ($E$) along the minimum dynamic path calculated with the MS-ARMD model into translational, rotational and vibrational kinetic energy as stated in equations \ref{etrans}-\ref{evib}.  The trajectory starts at the endo transition state and ends at the reactants.} 
		\label{classical-rot}
	\end{figure}

	\begin{figure}[H]
		\includegraphics[width=.5\textwidth]{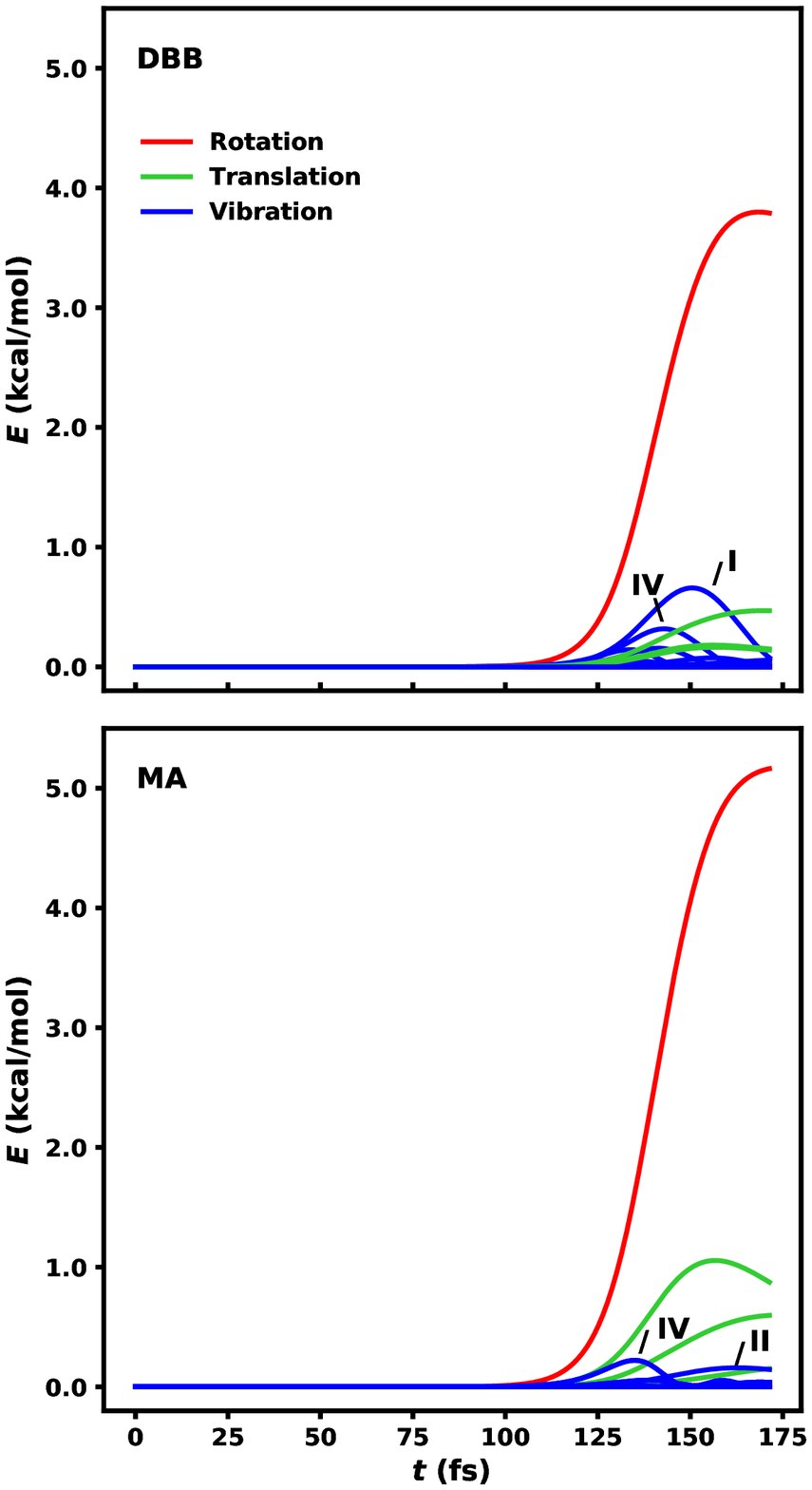}
		\caption{Projection of the total kinetic energy ($E$) along the minimum dynamic path calculated with the PhysNet neural network onto the degrees of freedom of dibromobutadiene (DBB) and maleic anhydride (MA). The trajectory starts at the endo transition state and ends at the reactants. The most important vibrations for DBB are (I) molecular out-of-plane bending and (IV) C=C in-plane symmetric bending. For MA they are (II) C=O out-of-plane symmetric bending and (IV) out-of-plane bending of the molecule.} 
		\label{NN-br-all}
	\end{figure}

	\begin{figure}[H]
		\includegraphics[width=.85\textwidth]{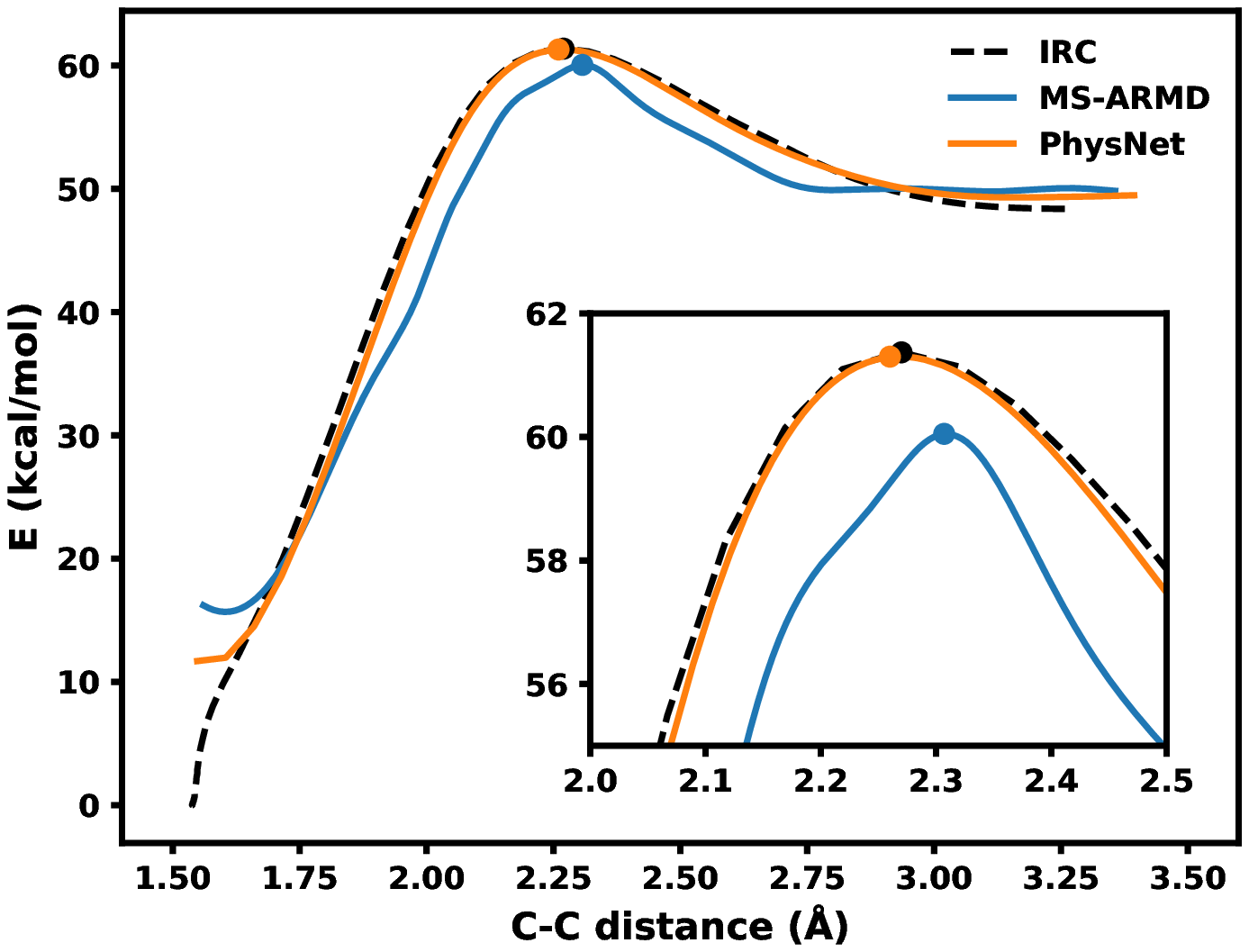}
		\caption{Comparison of the shape of the potential energy surface explored by the complete minimum dynamic path performed with MS-ARMD (in blue) and PhysNet (in orange) models as a function of the C-C distance of the bonds formed (the distances of the two bonds formed are identical along the trajectory). The internal reaction coordinate (IRC) calculated at the M06-2X/6-31G* level of theory is also shown as a function of the same coordinate in black. The position of the transition state is marked as a dot. The reactant state is at a C-C distance of 3.35~$\rm \AA$~and the product at a distance of 1.6~$\rm \AA$. A magnification of the transition state region is shown in the inset.} 
		\label{mdp-pes}
	\end{figure}
	\begin{figure}[H]
		\includegraphics[width=.5\textwidth]{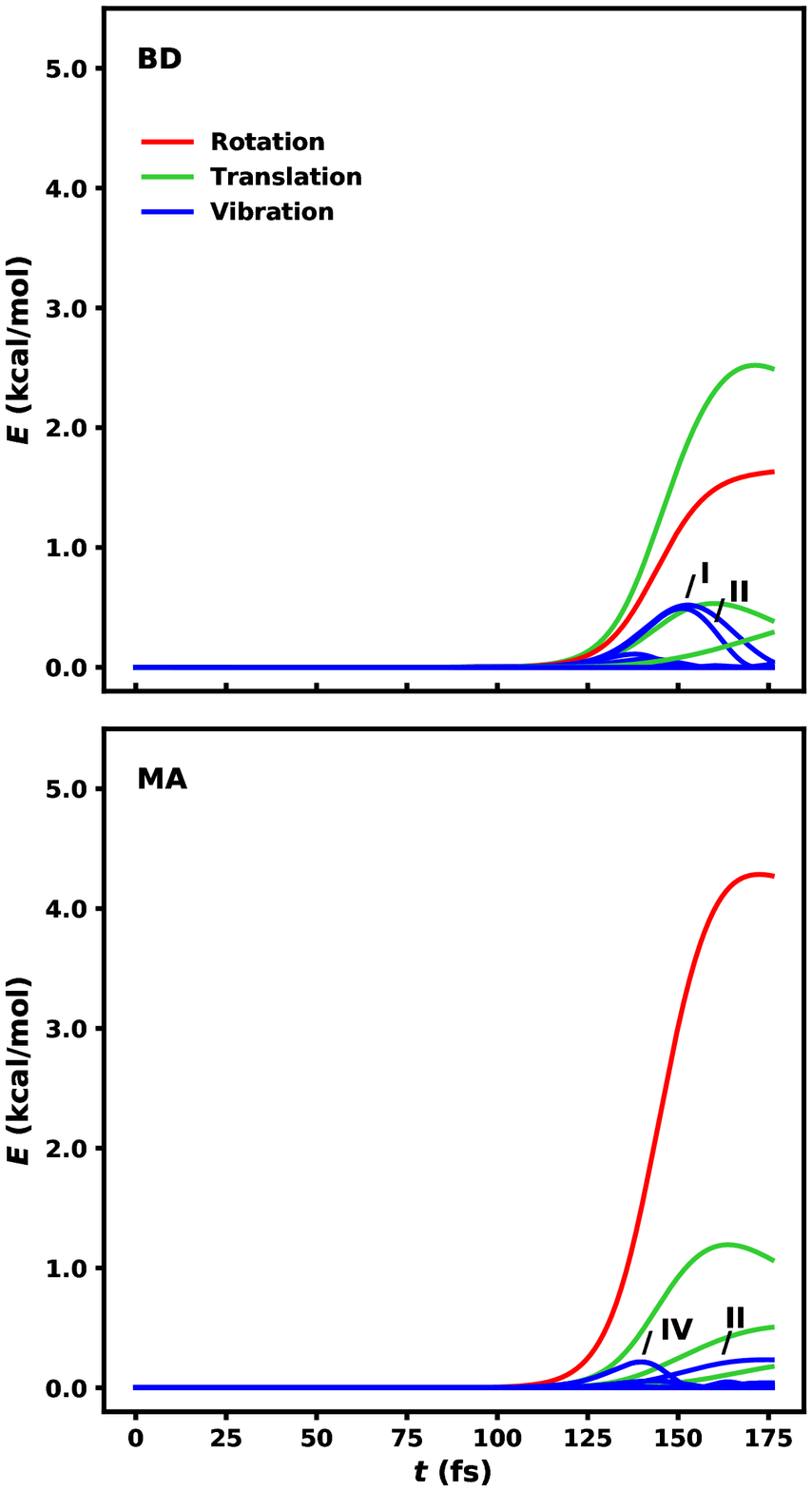}
		\caption{Projection of the total kinetic energy ($E$) along the minimum dynamic path calculated with the PhysNet neural network onto the degrees of freedom of butadiene (BD) and maleic anhydride (MA). The trajectory starts at the endo transition state and ends at the reactant. The most important vibrations for butadiene are (I) C=C in-plane symmetric bending and (II) C-H out-of-plane symmetric bending. For MA they are (II) C=O out-of-plane symmetric bending and (IV) out-of-plane bending of the molecule.} 
		\label{NN-h-all}
	\end{figure}

\section{Reactive events}
  \begin{table}[H]
    \caption{Initial conditions for the 482 recorded reactive events in terms of collision energy ($E_{\rm coll}$), impact parameter ($b$) rotational temperature ($T_{\rm rot}$) and vibrational temperature $(T_{\rm vib}$). $N$ is the total number of trajectories for each set of initial conditions.}
    \label{table-reac}
  \begin{tabular}{c|c|c|c|c|c}
  $E_{\rm coll}$ (kcal/mol) & ~~~~$b$ ($\rm \AA$) ~~~~& ~~~$T_{\rm rot}$ (K)~~~ & ~~~$T_{\rm vib}$ (K)~~~ & \# Reactive events & ~~~~~$N$~~~~~ \\  \hline 
  75 & 0 - 1 & 0 & 100 & 1 & 10$^5$ \\
  75 & 0 & 1000 & 100& 1 & 10$^5$ \\
  75 & 0 & 2000 & 100& 5 & 10$^5$ \\
  75 & 0 & 4000 & 100& 4 & 10$^5$ \\
  
  100 & 1 - 2 & 0 & 100& 1 & 10$^5$ \\
  100 & 0 & 1000 & 100& 6& 10$^5$  \\
  100 & 0 & 2000 & 100& 8 & 10$^5$ \\
  100 & 0 & 4000& 100 & 24 & 10$^5$ \\
  100 & 0 & 4000& 100 & 174 & 9$\cdot$10$^5$ \\
  100 & 0 - 1 & 4000 & 100& 129 & 10$^6$ \\
  100 & 1 - 2 & 4000 & 100& 49 & 10$^6$ \\
  100 & 2 - 3 & 4000 & 100& 39 & 10$^6$ \\ 
  100 & 3 - 4 & 4000 & 100& 32 & 10$^6$ \\
  100 & 4 - 5 & 4000 & 100& 3 & 10$^6$ \\
  \hline
  15 & 0 & 8000 & 100 & 1 & 10$^5$ \\
  20 & 0 & 8000 & 100 & 2 & 10$^5$ \\ \hline
  75 & 0 & 0 & 1000 & 1 & 10$^5$ \\
  100 & 0 & 0 & 1000& 2 & 10$^5$ \\  
  \end{tabular} 
  \end{table}

	\begin{figure}[H]
		\includegraphics[width=.45\textwidth]{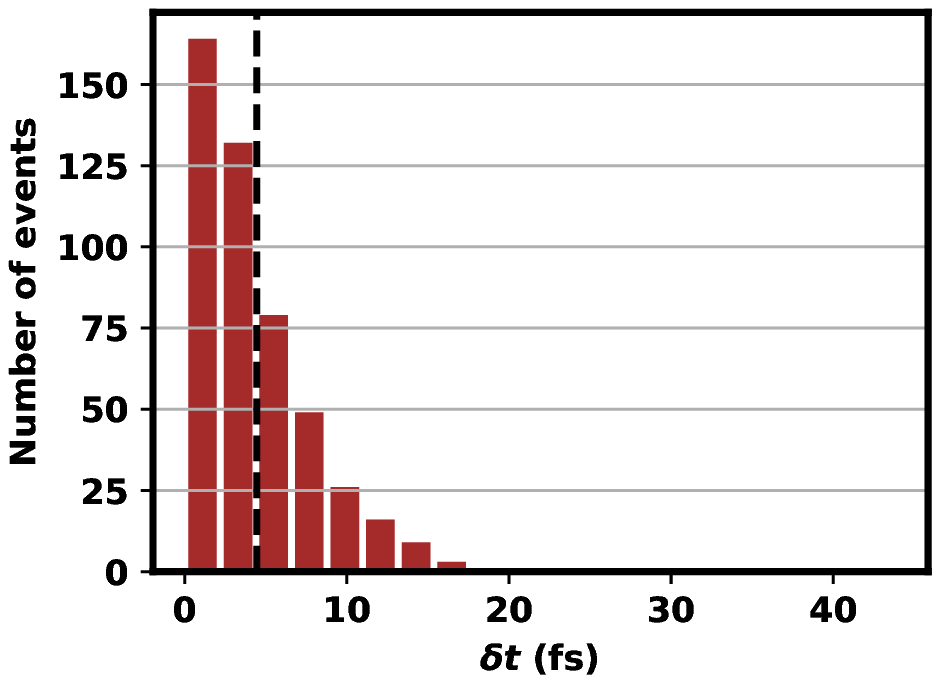}
		\includegraphics[width=.45\textwidth]{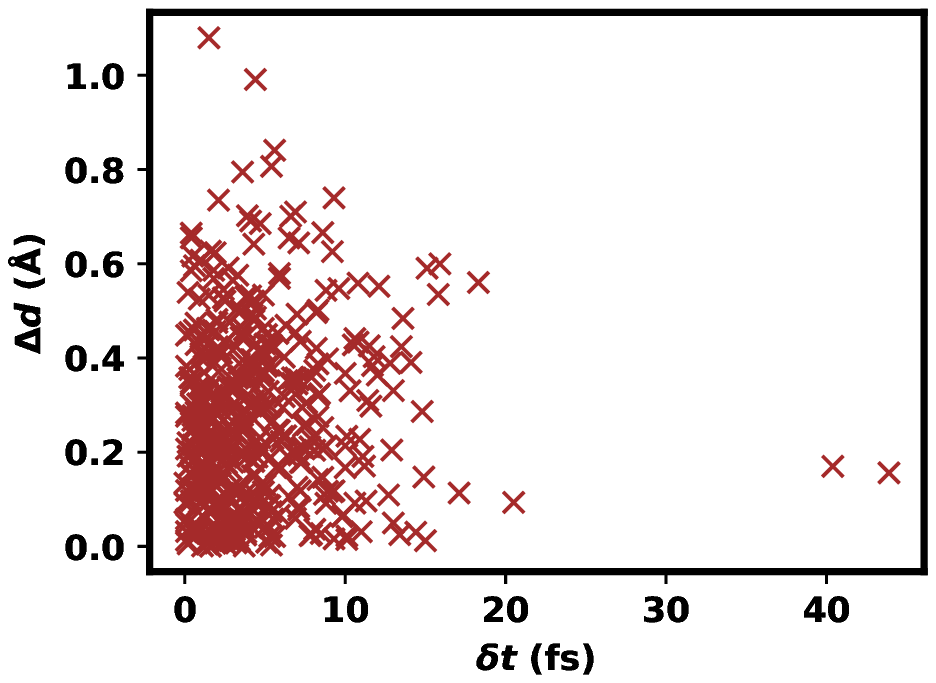}		
		\caption{Left panel: histogram of the elapsed time ($\delta t$) for all reactive events. The mean of the distribution appears as a vertical line. Right panel: difference in the C-C distances of the two forming bonds at the transition state structure ($\Delta d$) versus elapsed time of formation of the two bonds ($\delta t$).} 
		\label{synchronicity}
	\end{figure}